\pdfoutput=1
\documentclass[12pt, a4paper]{report}

\usepackage{booktabs}
\usepackage{amsmath, amsthm, amssymb}
\usepackage[sf,md,up,medium,raggedright,compact]{titlesec}

\usepackage{amsmath}
\usepackage{graphicx}
\usepackage{latexsym}
\usepackage{amsfonts}
\usepackage{txfonts}
\usepackage[sf,md,up,medium,raggedright,compact]{titlesec}

\usepackage[mediumspace,mediumqspace,Grey,squaren]{SIunits}

\usepackage{graphicx}

\usepackage{fancyhdr}
\usepackage[hmargin=1cm,vmargin=1.5cm]{geometry} 
\begin{titlepage}
\title{Integrated photonic 3D waveguide arrays for quantum random walks on a circle} 
\author{Trond Linjordet \vspace{1cm} \\
Supervisors: Prof. Jason Twamley, Dr. Graham Marshall, Assoc. Prof. Stojan Rebic \\
Department of Physics and Engineering, Macquarie University, North Ryde, Sydney \vspace{3cm} \\}
\date{December 7, 2009 \vspace{1cm} \\
Submitted in partial fulfilment of the requirements for the Honours\\
Degree of Bachelor of Science (Physics) in the Faculty of Science\\
at Macquarie University, 2009.}
\end{titlepage}

\begin{document}

\maketitle

\begin{abstract}
\vspace{1cm}
\begin{center}
\end{center}
\vspace{1cm}
Quantum random walks (QRWs) can be used to perform both quantum simulations and quantum algorithms. In order to exploit this potential, quantum walks on different types of graphs must be physically implemented. To this end this we design, model and experimentally fabricate, using the femtosecond laser direct-write technique, a 3D tubular waveguide array within glass to implement a photonic quantum walk on a circle. The boundary conditions of a QRW on a circle naturally suggests a 3D waveguide implementation - allowing much simpler device design than what could be achieved using a 2D waveguide architecture. We show that, in some cases, three-dimensional photonic circuits can be more suited to the simulation of complex quantum phenomena. 
\end{abstract}
\newpage

\chapter*{Contributions and Acknowledgements}

I want to express my gratitude to my supervisors Jason Twamley, Graham Marshall and Stojan Rebic for their patience, kindness, support and guidance in this project. Thanks to Jonathan Matthews of Bristol University, who wrote the Mathematica scattering QRW code, which generated  Fig.~\ref{ScatteringQRWplay}. I also want to express my gratitude to the department in general, as several figures used in this thesis were kindly provided by members of the Department of Physics and Engineering. \\
The fundamental concept for this project came from my supervisors and arose over the course of several conversations I had with them. The tubular waveguide array concept came up in a discussion between myself, my supervisors and Jonathan Matthews. I subsequently worked with these ideas and made some simulations in two different programs, comparing various configurations of design parameters for the tubular array. Ultimately, I came to what I felt was an optimal compromise between the various constraints for fabrication and the intended function of the tubular array. \\
As I have not been trained in the femto-second laser direct-write technique and other crucial techniques for these experiments, I could not be primarily responsible for the fabrication and testing of the tubular waveguide array devices. However, while Graham Marshall performed most of the work in device fabrication and characterisation, I assisted at each step experimentally. I also made the decision to spend more time fabricating fewer devices, as this could possibly allow us to explore higher refractive index waveguides. This gamble appears to have paid off, but was risky since it increased the possibility of fabrication errors.

\tableofcontents

\chapter{Introduction}

The modern world depends on information technology, and since 1962 computing power has roughly doubled every two years. Here, computing power is conceptualized as the maximum number of transistors that can fit in a given area of an inexpensive integrated circuit \cite{Moore1965, Thompson2006, Moore1975}. Society continuously demands increased computing power in the military, industry, business, medicine, scientific research and for private purposes. This has lead to continuous cycles of development and obsolescence in information technology. Hence the progress of information processing power is important both to the infrastructure and the economy of the global society. Effectively, this depends on making processing power denser. However, the standard method of producing very fine integrated circuitry using UV light lithography is progressing closer to its inherent physical limits of miniaturisation \cite{Powell2008}. What will be the next step in pushing the limits of information processing hardware? As miniaturisation and more innovative approaches progress, information processing will eventually involve device components of atomic or molecular scale. Thus, whether by conventional semiconductor transistors or other technologies, the physical role of quantum coherence will become more important in information technology. Accordingly, quantum mechanics will increasingly be required to describe, design and understand information processing hardware in the future. 
\\

Quantum computing is different from conventional computing, because it is subject to different logic and physical constraints. Hence it requires a different theoretical understanding and precise engineering \cite{NielsenChuang}. However, quantum computing holds the promise of significant advantages over conventional computing.
\\

Quantum mechanics describes the world of atomic and subatomic particles, revealing a set of laws that differ fundamentally from the classical laws of the macroscopic world. By understanding the laws of quantum physics and using refined instruments, various technologies have been developed that harness the effects of the quantum mechanical world. Nuclear power, lasers, electron microscopy, nuclear magnetic resonance imaging are just a few examples of technologies which were developed through applications of quantum mechanics. 
\\

As the available instrumentation is being increasingly refined, much attention and effort is being devoted to developing computers based on quantum effects, in the hopes that systems can be developed that outperform conventional computers, at least in specific tasks. For some of these tasks, theoretical algorithms already exist. These tasks which classical computing would perform in \textit{exponential} time, quantum computing can accomplish in only \textit{polynomial time}. In other words, for a problem of size $N$, i.e. where $N$ items must be treated, a classical computer might need to spend computing time on the order of $e^N$ while the quantum computer could do the task in computing time as some polynomial of $N$. For example, the Shor quantum algorithm efficiently factors numbers into component primes, and can thus break any classical encryption which is based on the multiplication of large prime numbers  \cite{Shor1994}. Efficiently here means that the computational resources merely grow polynomially, not exponentially, relative to the size of the problem. A search algorithm for unsorted databases has also been developed which functions with quadratic speed-up compared to the best known classical search algorithms, which scales linearly with the size of the database \cite{Grover1996}. The Shor algorithm has primarily military, intelligence and security applications, but any endeavour that avails itself of information technology will deal with databases. The possibility has also been postulated to use quantum computing to simulate quantum systems more efficiently than a conventional computer \cite{Feynman1982, Gerritsma2009}, and this could benefit all the physical sciences. This provides a strong motivation to investigate the potential hardware and software of quantum computing devices.  
\\

How can a quantum computer be implemented? One of the primary challenges is to protect a quantum system from environmental noise, which would cause computing errors if quantum states are unintentionally decohered. To make quantum computing work, a quantum computer must be built which protects the quantum information from noise. Also, another challenge is communicating between the different realms, i.e. between the classical physics and quantum physics portions of a device. In a quantum system, reading data must involve taking a measurement of the system. This in turn changes the quantum state measured. For this reason, the logic of quantum computing differs from the logic of conventional computing, where reading data does not change the stored data. A qubit is the quantum information equivalent of the classical computer bit, i.e. it is a two-level quantum system used to store or express data. Entanglement is a quantum physical phenomenon of non-classical, non-local correlation of different parts of a quantum system. This phenomenon is an essential part of quantum computing and quantum information, although the reasons why are not yet properly understood \cite{NielsenChuang}. Experiments have demonstrated that quantum computing can be implemented in principle \cite{Walther2005, Vandersypen2001}. However, the challenge remains to find an appropriate, robust form of implementation that can be scaled up to solve more complicated problems. In other words, assembling and coordinating quantum computing resources to deal with problems which are not trivial in conventional computing. Only when this is achieved can quantum computing become a useful technology. Further developments can bring quantum computing to its full potential, surpassing the capabilities of conventional computing in specific areas.
\\

There are many different kinds of implementations of quantum computing which are currently being investigated. These different approaches all base themselves on different technologies to encode, control and measure quantum information physically. A few different approaches that are being investigated include nuclear magnetic resonance (NMR), superconductor chips, ion traps and optical implementations \cite{NielsenChuang, negrevergne:170501, Dicarlo2009, Haffner2005} . We note that in almost any practical implementation, photons play a pivotal role in measuring, manipulating and transferring quantum information. Recently, implementing quantum computing optically with integrated quantum photonics has received interest due to the potential stability and scalability of this technology \cite{Rieffel2008}.  This project is based on such an optical implementation. 
\\

More specifically, this project uses photonic quantum science to investigate quantum random walks (QRWs). Classical random walks consist of a series of steps, where for each step there is more than one direction to take the next step, with a certain probability associated with each direction. The simplest case can be imagined as a person walking on a line and flipping a coin before each step to decide whether the next step will go forwards or backwards. This literal example can be translated into a position on an integer number line that changes by one for each step, with a 50 \% chance of going in each direction. The mathematics of random walks have been involved in conventional computing algorithms, as randomness has been found to make many algorithms more efficient or to simplify them relative to their deterministic equivalents \cite{Mitzenmacher2005}. 
\\

To understand the difference between classical random walks and QRWs, it is important to understand the role of quantum superposition. Quantum superposition means that a coherent quantum system can exist in many different possible states or positions simultaneously until a measurement is made, causing the system to decohere and collapse into a single position or state. QRWs are similar to classical random walks, but using the principle of quantum superposition, whereby instead of choosing one path out of many possible, the quantum random walker walks all possible paths simultaneously. It is important to note that the QRW is therefore not in fact random --- and is sometimes referred to only as a quantum walk --- but if the quantum walker is decohered, the system immediately collapses to a single position, making it a classical random walk.  This can be used to speed up quantum algorithms relative to their conventional counterparts \cite{Childs2003}. It's important to note that on the one hand, many schemes have been proposed to physically implement QRWs,  e.g. with quantum dots and solid state charge qubits \cite{Manouchehri2008, Manouchehri2009}, or with single photonic qubits \cite{Zou2006}. On the other hand, these papers do not refer to actual experiments. Relatively few papers present actual experiments with QRWs, although a two-qubit NMR implementation of QRW has been reported by Du \textit{et al.}  \cite{du2003} and Ryan \textit{et al.} \cite{Ryan2005} have reported an eight step QRW on a three-qubit NMR implementation. Since then, Ribeiro  \textit{et al.}  \cite{ribeiro2008} have implemented a single step of a single-photon QRW.  
\\

It is far too ambitious to produce a device which runs an algorithm based on a QRW. Instead, this project investigates the physics of QRWs in a specific physical structure. The purpose of the project is to investigate the physics of the QRW in a waveguide array to create a level of understanding that enables future developments towards quantum computing. 
\\

\section {Aim and scope of project}

{\bf The aim and scope of the project can be summarised as an investigation of QRWs in an integrated quantum optical device.} An optical implementation of quantum computing depends on precise and reliable manipulation of photonic qubits. A array of adjacent waveguides hold the promise of a tool that can provide such control of quantum states by photon coupling between waveguides \cite{Bromberg2009}. Furthermore, waveguides can be fabricated in an integrated architecture to perform operations equivalent to bulk optics devices in a far more compact space \cite{OBrien2007, Politi2008}. 
\\

\begin{figure}[hbt]
  \begin{center}
    \resizebox{!}{60mm}{\includegraphics{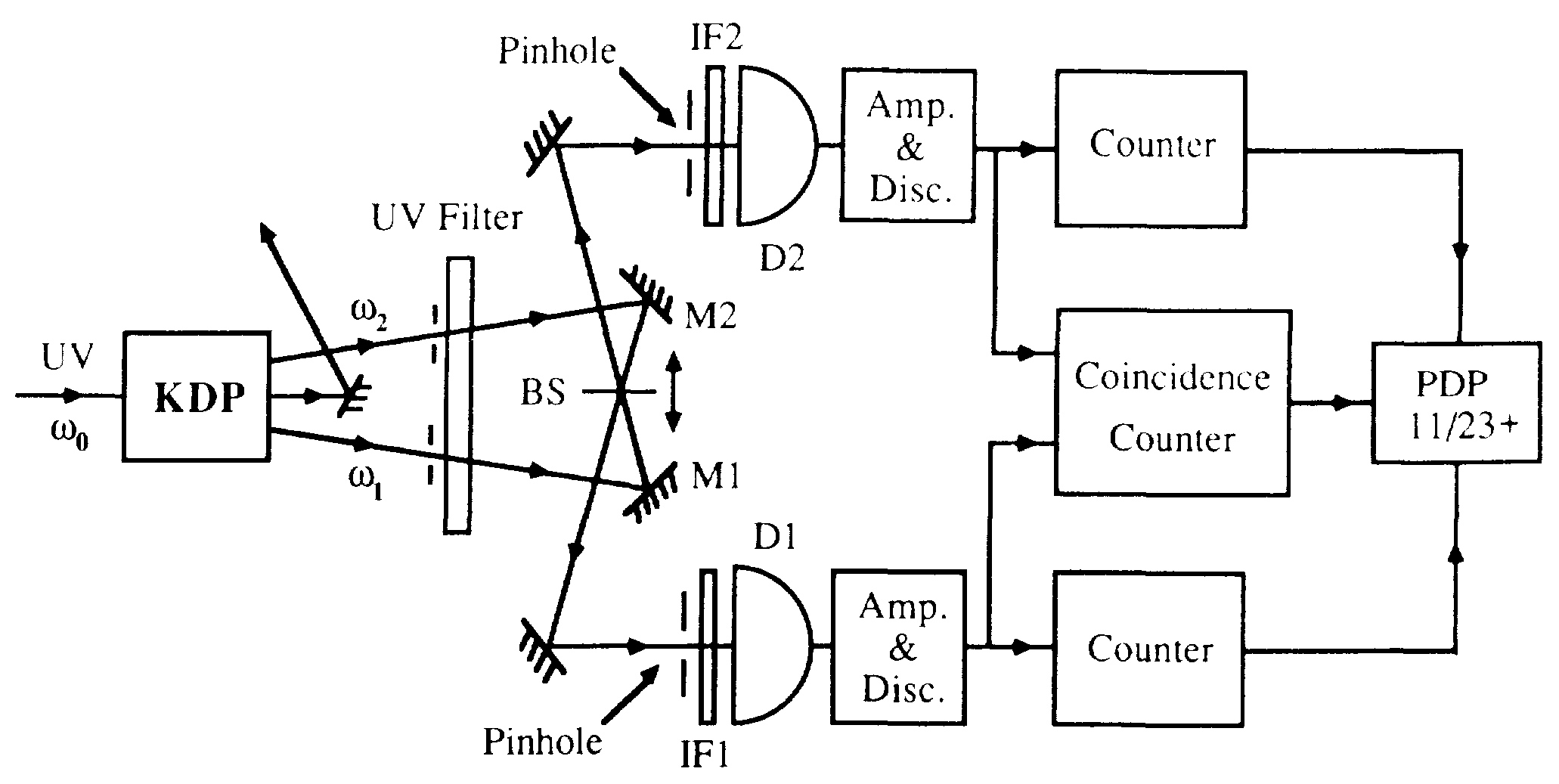}}
  \end{center}
  \caption{Experimental setup to observe path entanglement. (Reproduced from \cite{HOM1987}.) }
  \label{HOMsetup}   % <-- IMPORTANT: \label is *after* \caption
\end{figure}

Recently, Bromberg, \textit{et al.} \cite{Bromberg2009} investigated photonic QRWs in a planar waveguide array. Using simulations, they study the quantum mechanical interference of photon pairs propagating in planar waveguide arrays. This is done to see how the positions of the photons correlate after propagating through the waveguide array. Analogous experiments with classical bright light are performed where possible, and these correlation matrices correspond to those of the quantum mechanical simulations, but with reduced contrast. No quantum mechanical experiments are performed, all the experimentation is done with classical light. Light is injected at one side of the array and the correlation matrices show how the exit positions of photon pairs are associated depending on the injection positions of the photons, and whether the photon pairs are entangled. Light is injected into either a single waveguide, two adjacent waveguides or next-to-adjacent waveguides. Both entangled and unentangled photon pairs are simulated, but here entangled photons are not both injected into the same waveguide. The entanglement employed is path entanglement, which means that the paths taken by the two photons are correlated non-classically. Photon path entanglement is observed by sending both photons into the same beamsplitter, at the same time, from different sides, and then allowing the photons to travel the same distance before entering either of two detectors. Given the described conditions, the two photons will both go to either of the two detectors. The experimental setup that was used to discover this effect is shown in Fig.~\ref{HOMsetup} \cite{HOM1987}. The Bromberg study has produced matrices correlating the exit positions of the two photons. These matrices indicate bunching and anti-bunching of photons due to quantum mechanical interference, and this is called the the Hanbury Brown-Twiss effect \cite{Bromberg2009}. The current project investigates this same topic but in a more ambitious manner, by studying a three-dimensional array of tubular waveguides. This three-dimensional structure significantly changes the pattern of coupling. The tubular structure has no boundaries, which would reverse the direction of the walk's propagation. The current project involves designing a platform to study three-dimensional QRW structures. No work published to date has considered a three-dimensional QRW experimentally. 
\\

\begin{figure}[hbt]
  \begin{center}
    \resizebox{!}{60mm}{\includegraphics{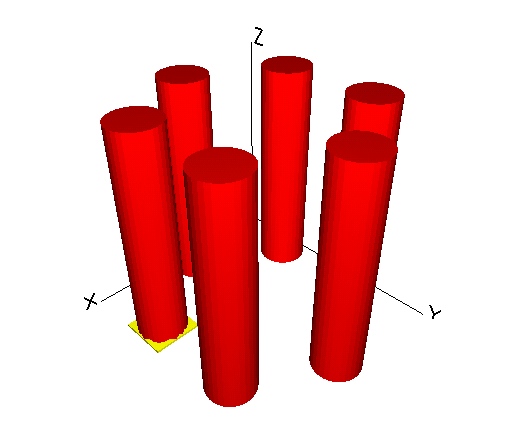}}
  \end{center}
  \caption{Image of a six-waveguide tube from the RSoft classical light simulator.}
  \label{6wgRsoft}   % <-- IMPORTANT: \label is *after* \caption
\end{figure}

This project creates integrated photonic waveguide networks
using the laser direct-write fabrication technique \cite{Marshall2009}. The aim is to explore the effect of a tubular waveguide array on bright classical light input. This architecture exploits the unique
three-dimensional advantage which the direct-write fabrication
technique has over the lithographic approach. In other words, this
structure could not be easily fabricated with any other technique. The
direct-write fabrication technique not only enables the fabrication of
the 3D structure, it also is much quicker, requiring less processing
time with fewer steps of fabrication than a comparable length of
waveguide created with lithography. Six parallel waveguides are
created in a tubular array, as shown in Fig.~\ref{6wgRsoft}. The integrated waveguide circuits that are
fabricated have a coupling network equivalent to approximately 96 bulk optics beam splitters but are compressed to span a physical distance of merely 20-22 mm. By comparison, previously a photonic device (i.e. a CNOT gate) was lithographically created which consisted of four waveguides and with the coupling network equivalent to four bulk optics beamsplitters \cite{Politi2008}.
\\

\begin{figure}[hbt]
  \begin{center}
    \resizebox{!}{60mm}{\includegraphics{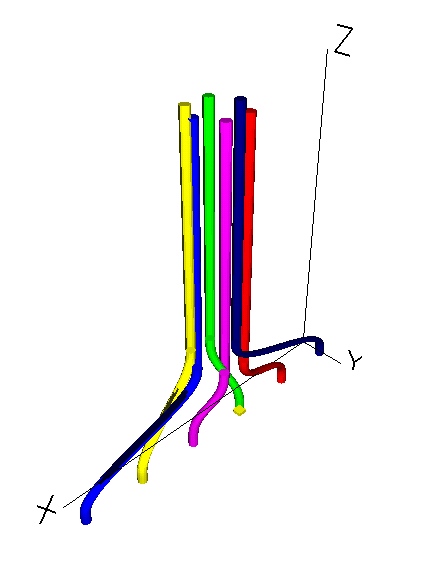}}
  \end{center}
  \caption{Image of a six-waveguide tube with a single stage fan-in section, from the RSoft classical light simulator.}
  \label{6wg1stageRsoft}   % <-- IMPORTANT: \label is *after* \caption
\end{figure}

\begin{figure}[hbt]
  \begin{center}
    \resizebox{!}{80mm}{\includegraphics{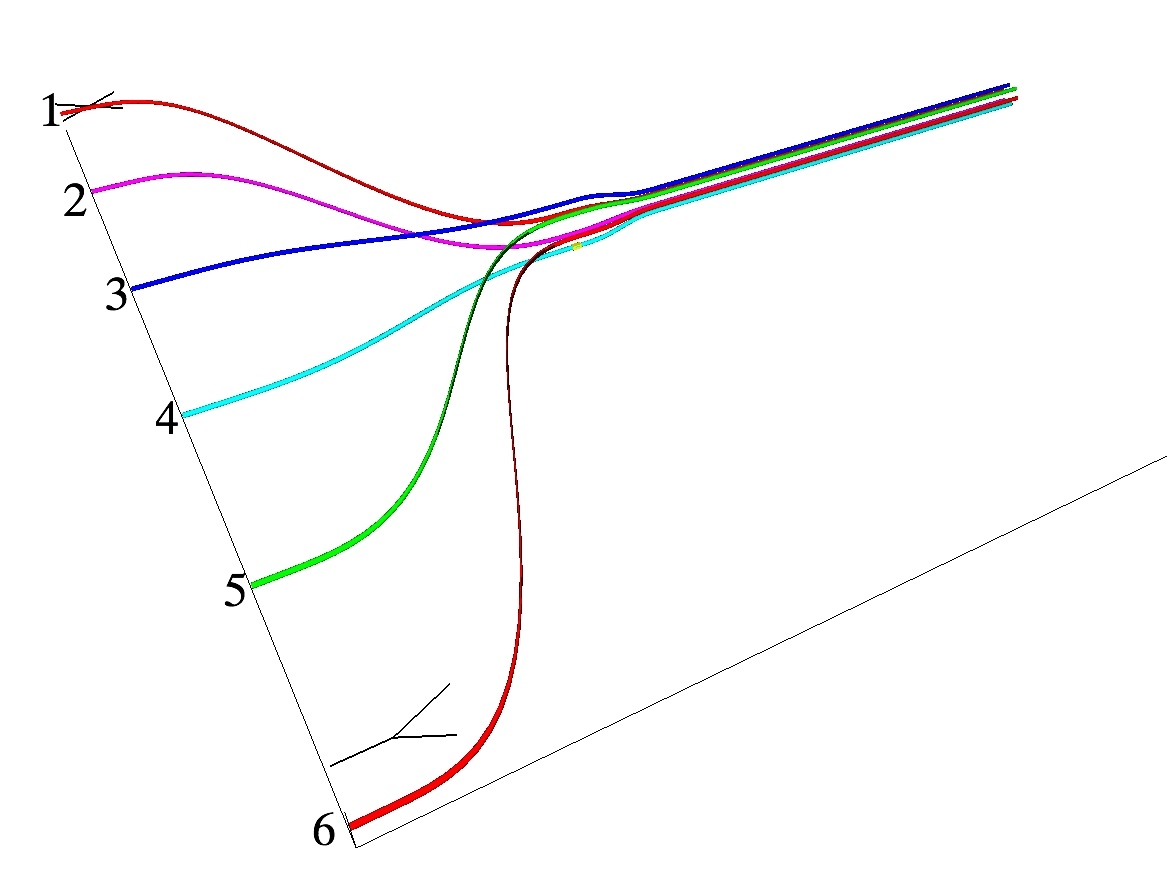}}
  \end{center}
  \caption{Image of a six-waveguide tube with a two-stage fan-in section, from the RSoft classical light simulator. The second stage is already in a tube form but has a larger radius than the interaction length of the tube. The numbers 1-6 indicate the numbering convention used throughout this project to keep track of which waveguides fan into which part of the tube.}
  \label{6wg2stageRsoft}   % <-- IMPORTANT: \label is *after* \caption
\end{figure}

As we will detail later in this thesis: 
At the end of this project, almost all of the initial goals have been achieved. The tubular array devices have been designed and fabricated, and the resulting two six-waveguide devices have been measured with beam profilometry and refractive contrast profilometry. These two devices were designed differently with respect to the fan-in section leading light from bulk optics light sources into the tube while minimising interference and cross talk between waveguides before they reach the tube. The first device has a single fan-in stage, while the second has two stages in
the fan-in section, as shown in Fig.~\ref{6wg1stageRsoft} and Fig.~\ref{6wg2stageRsoft}, respectively. The simulations have also been made up to the point where the quantum mechanical case can begin to be considered in detail. Quantum mechanical simulations involving a single photon were performed by taking classical bright light to be equivalent to a single photon. However, due to time constraints, completely quantum mechanical simulations for multiple photons as well as quantum mechanical experiments are pending. With regard to reproducing the results of Bromberg \textit{et al.}, classical bright light measurements have been made, shining 780 nm light into each waveguide and recording the resulting beam profilometry at the exit of the tubular waveguide arrays. This novel device appears to have two different tubular waveguide arrays with no fabrication errors. Hence, the tubular waveguide arrays are ready to be experimentally tested with pairs of real single photons, both entangled and unentangled. In the latter case we expect little difference in correlation matrices from classical bright light. 
\\

For the system of interest, the waveguide array without boundaries, the fabrication results are promising, but some theoretical concerns remain. Experimentally ideal conditions could not be met in terms of refractive contrast and waveguide width, but the way to physically fabricate an optimal system has been investigated and the challenges have been identified and theoretical solutions have been proposed. A more radical re-design is also being considered, as outlined in the conclusion. The project builds on the results of Bromberg, \textit{et al.} \cite{Bromberg2009} .  \\

In Chapter 2, I review literature relevant to the fields which the project falls within, as well as literature relevant to the project itself.\\

In Chapter 3, the theory behind the project is presented and the process of simulating and designing the device is explained. Simulation results are included in the explanation. \\

In Chapter 4, the methods used to fabricate and measure the device are explicated, and the experimental results are shown. \\

In Chapter 5, the theoretical and experimental results are discussed. \\

%**********************************************************************
%
%														Chapter 2 - Literature review
%
%**********************************************************************

\chapter{Literature review}

\section{Overview}

With reference to literature, explanations are given about what QRWs are and how they can benefit quantum computing. The next topic how optics can be used to physically realise quantum information designs. Furthermore, what are the advantages of doing so with integral photonics? Also, the laser direct-write fabrication technique, the specific fabrication technology in this project, is presented. Finally, we look at relevant previous work in the intersection of QRWs and optical realisations. 

\section{Quantum random walks}

\begin{figure}[hbt]
  \begin{center}
    \resizebox{!}{30mm}{\includegraphics{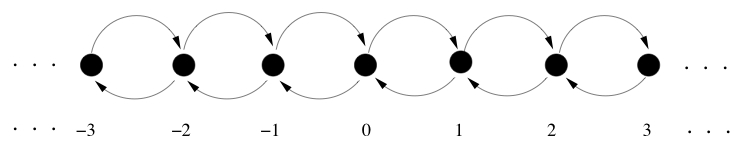}}
  \end{center}
  \caption{A classical random walk on an infinite line. (Reproduced from \cite{Venegas2005}.) }
  \label{Classical random walk}   % <-- IMPORTANT: \label is *after* \caption
\end{figure}

To define a random walk, one can first consider the simplest, classical case. Given a line divided into discrete steps, with a walker on the line --- often described as a drunkard --- a coin toss with an unbiased coin can be used to decide, for each step, whether the walker will take a step forward or backward along the line. Taking the line to be a set of integers and the walker begins at zero, then after $t$ discrete time steps, the walker will be somewhere along the number line between $-t$ and $+t$, as illustrated in Fig.~\ref{Classical random walk}. \\

\begin{figure}[hbt]
  \begin{center}
    \resizebox{!}{60mm}{\includegraphics{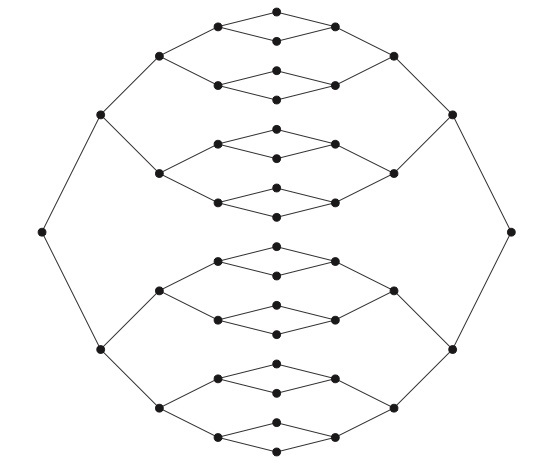}}
  \end{center}
  \caption{The graph $G_4$, or glued binary tree. We can label the columns of vertices in this type of graph. $G_n$ has $2n+1$ columns of vertices. (Reproduced from \cite{Childs2002}.) }
  \label{G4}   % <-- IMPORTANT: \label is *after* \caption
\end{figure}

If a number of classical random walks occur on the line, each starting at position zero and each evolving for a time $t$, and record the final position of each random walk, then the distribution of final positions approaches the Gaussian distribution as the number of random walks increases. The variance is $\sigma^2 = t$ and consequently standard deviation is $\sigma = \sqrt{t}$ \cite{Inui2005, Kempe2003}. More complicated random walks occur when the walker is placed on a graph --- a set of vertices  interconnected by edges --- which is not the number line. Take for example Fig.~\ref{G4}, the graph $G_4$. \\

\begin{figure}[hbt]
  \begin{center}
    \resizebox{!}{90mm}{\includegraphics{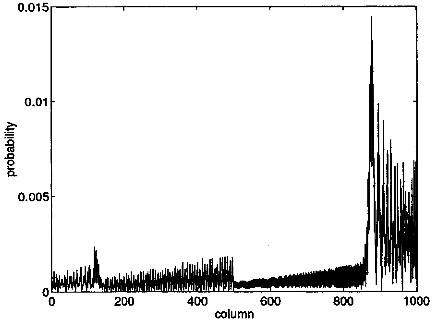}}
  \end{center}
  \caption{Probability distribution of the quantum walker being in each of the $2n+1$ columns in $G_n$ where $n=500$, after $400$ time steps. The walker has already walked through the entire graph because the quantum walk propagates with speed $2\sqrt{2}$. (Reproduced from \cite{Childs2002}.) }
  \label{G4walk}   % <-- IMPORTANT: \label is *after* \caption
\end{figure}

In this type of graph, $G_n$, two binary trees of depth $n$ are ``glued together'' as shown in Fig.~\ref{G4}. We start a classical random walker at the leftmost vertex and then ask for the average time for that walker to traverse to the rightmost vertex. In the classical random walk the probability of reaching halfway across this graph is large. However, from the halfway point, the probability to step farther decays exponentially. One can estimate that the classical random walker has a probability less than $2^{-n}$ of traversing $G_n$ irrespective of how long they walk \cite{Childs2002}. By comparison, the QRW propagates across the graph linearly with $n$. In Fig.~\ref{G4walk}, after just 400 steps the leading edge of the QRW probability distribution has already reached and been reflected back from the rightmost root vertex of $G_{500}$ \cite{Childs2002}. \\

In theoretical computer science, the classical random walk constitutes a fundamental concept and is used in various algorithms, such as Monte Carlo applications \cite{Papa1994}. Some of these algorithms have a significant probability of error, but can be repeated many times to get an arbitrary degree of probabilistic accuracy. \\

\subsection{Coined quantum random walks}

There are different types of quantum random walks. The two main models are the \textit{discrete time} and the \textit{continuous time} quantum random walks. Both are being explored as possible means of constructing novel or significantly more efficient quantum algorithms \cite{Kempe2003}. The coined QRW is a kind of discrete QRW which not the subject of this project. Nevertheless, the coined QRW illustrates the concept of QRWs and is a simpler case than the continuous QRW, which this project has dealt with. Also, discrete QRWs may become relevant in future work, as will be discussed later.  Historically, the discrete QRW also precedes the continuous QRW, since the discrete QRW was the form of quantum random walk which was first explored in the work of Aharonov \cite{Aharanov1993}.  \\ 

An important concept in quantum mechanics is Hilbert space. The Hilbert space is a complex vector space with a well-defined inner product. The vectors in this space are used to represent quantum states \cite{NielsenChuang}.
\\
The discrete QRW is similar to the classical random walk in that time progresses by discrete intervals. The discrete QRW is described by a Hilbert space $\mathcal{H}$ consisting of the graph and a complex space representing the quantum "coin." For example, for the quantum random walker on the integer number line, we have $\mathcal{H}=\mathbb{Z} \otimes \mathbb{C}^2$. Here $\mathbb{Z}$ represents the integer line, and $\mathbb{C}^2$ represents the two level quantum coin.\\ 

\begin{figure}[hbt]
  \begin{center}
    \resizebox{!}{60mm}{\includegraphics{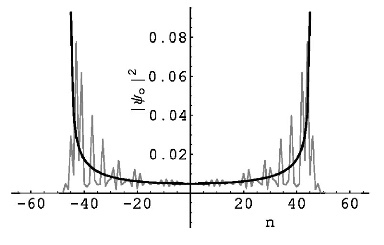}}
  \end{center}
  \caption{Probability distribution of the symmetric quantum random walk after 32 discrete time steps, showing a wavelike symmetric propagation away from the origin. The axis labelled $n$ is the position axis. (Reproduced from \cite{Patel2005}.) }
  \label{Hypercube distribution}   % <-- IMPORTANT: \label is *after* \caption
\end{figure}

First, due to the superposition principle, a quantum walker will not simply traverse one particular trajectory. A single quantum random walk consists of all possible walks consistent with the starting state, in superposition. Thus QRWs are not really random per se. Furthermore, intermediate steps or iterations interfere quantum mechanically with each other. For example, the symmetric QRW on the number line will have a probability distribution that differs markedly from the Gaussian and has two peaks away from the centre \cite{Patel2005}. This is illustrated in Fig.~\ref{Hypercube distribution}. Here, the QRW, when viewed on the walker's line, is symmetric because the initial state is taken to be \mbox{$| \Phi_{sym} \rangle = \frac{1}{\sqrt{2}}(|\uparrow\rangle + i|\downarrow\rangle)_C \otimes |0\rangle_W$} , where $| \downarrow \rangle _C$ is the state of the quantum coin and $| 0 \rangle _W$ is the state when the walker is completely localised at the origin. This evolution results in a probability distribution to find the quantum walker at a specific location on the line which peaks far away from its initial position.

Since while propagating the QRW we don't wish to interfere with, measure or decohere the quantum system which is highly susceptible to errors and noise, the equivalent of a classical coin flip is not applied at each iteration of a QRW. Instead, we then ``flip'' the coin by applying a unitary operation called the Hadamard gate $\hat{H}$, to the ``coin'' state,  \mbox{$\hat{H}=|j\rangle_C \rightarrow \frac{1}{\sqrt{2}} (|0 \rangle_C +(-1)^j|1 \rangle_C)$}, and then a conditional step depends on the state of the coin, $\hat{S}$: \\
 
\mbox{$\hat{S}=| \downarrow \rangle _C | j \rangle _W \rightarrow | \downarrow \rangle _C | j+1 \rangle _W$}
\\

\mbox{$\hat{S}=| \uparrow \rangle _C | j \rangle _W \rightarrow | \uparrow \rangle _C | j-1 \rangle _W$}\\

% 
% | \Phi_{step1} \rangle = \frac{1}{2}(|\uparrow \rangle_C \otimes (|1\rangle_W) + i|\downarrow\rangle_C\otimes (|-1\rangle_W))

It has the effect equivalent to a coin flip, although no outside randomness is introduced into the quantum system. \\

\subsection{Continuous quantum random walks}

This project focused on a physical system where injected photons perform a continuous QRW because of the geometry of the waveguide array being studied. As the photon travels through one waveguide, there is a continuous probability that it couples to another waveguide. Therefore, the continuous QRW is an integral concept in this project.\\

The continuous time model of quantum random walks was developed by Farhi \textit{et al.} \cite{Farhi1998} based on the idea of Markov chain processes. The continuous QRW does not have a ``quantum coin.'' The entire graph is represented by the time independent Hamiltonian $\hat{H}$ of the system, a matrix with non-zero off-diagonal elements. Here each non-zero element represents the edge between two vertices in the graph and the transition rate of moving from one vertex to the other. However, to make the process continuous, the walker can step between any adjacent connected vertices at any time. The walker does so with jumping rate $\gamma$ per unit time. \\

Thus the off-diagonal elements corresponding to connected pairs of vertices are equal to $-\gamma$ while the elements on the diagonal are equal to $k \gamma$ where k is the number of edges connected to the vertex which is represented by the diagonal element. The negative sign in the off-diagonal elements represents probability of transitioning away from the vertex in question, while the positive  elements in the diagonal represent probability of transitioning into the relevant vertex. Both the continuous and discrete QRWs propagate in linear time on the graphs $G_n$. \\   

\begin{figure}[hbt]
  \begin{center}
    \resizebox{!}{60mm}{\includegraphics{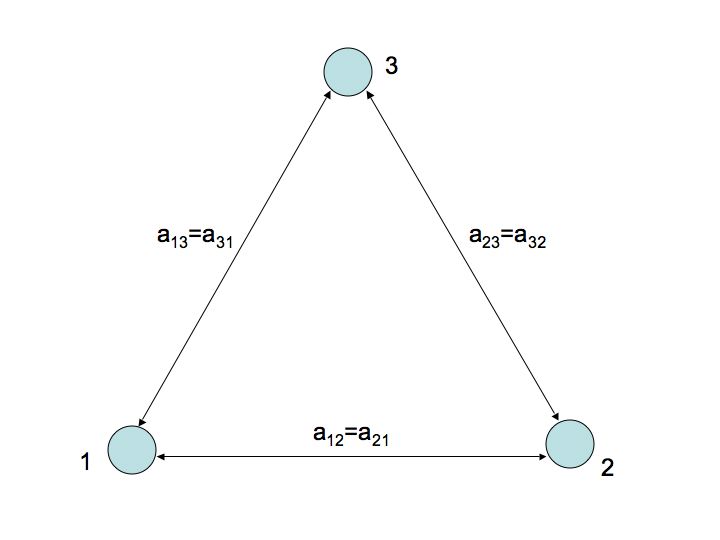}}
  \end{center}
  \caption{A three vertex circle with some transition probability rate elements shown. }
  \label{simpleContQRW}   % <-- IMPORTANT: \label is *after* \caption
\end{figure}

We can illustrate the continuous QRW in a simple, relevant form. Given the continuous QRW on a three-step circle, as in Fig.~\ref{simpleContQRW}, transitions between vertices occur with a probability $\gamma$ per unit time. Hence, the walk can be described by a Hamiltonian matrix  $\hat{H}$, where each element represents the transition probability per unit time. Each element $a_{ij}$ represents the transition probability rate from element i to j. The Hamiltonian is expressed as

 $\hat{H}= \begin{bmatrix} 2\gamma & -\gamma & -\gamma \\ -\gamma & 2\gamma & -\gamma \\ -\gamma & -\gamma & 2\gamma  \end{bmatrix} $.
 
 We then have that the unitary time operator 

 $\hat{U}(t)=\exp{}(-i \hat{H} t )$. 

For a more general and detailed explanation, see Kempe \cite{Kempe2003} and Farhi \cite{Farhi1998}.

\subsection{Useful properties of quantum random walks}

Quantum random walks can behave in surprising ways, but there are some quantitative properties which can be used to compare different types of QRWs. Two important concepts in characterising different types of QRW are the \textit{mixing time} and \textit{hitting time} \cite{Kempe2003}. Essentially, the mixing time is the time a particular QRW takes before taking on some characteristic stationary distribution, so that successive steps do not significantly alter the shape of the distribution. Some QRWs never reach a stationary distribution, continuing to propagate as long as the walk continues. These can not be said to have a mixing time. The hitting time is the amount of time required for a walk to reach a specific point, such as a particular vertex on the graph being walked.

Another important property is the position variance of the final position of the walker. The variance of final position for the simple QRW $\sim t^2$, hence the standard deviation or expected distance from the starting point $\sim t$ \cite{Ambainis2001}. Therefore the quantum random walk expands linearly with time, quadratically faster than the classical random walk. The linear expansion means that quantum random walks may potentially perform algorithms much more efficiently than classical random walks. \\

Specifically, in the case of a random walk over a circle (a graph of $N$ vertices connected in a circle or in the shape of a regular polygon), the mixing time of a quantum random walk is likewise quadratically shorter than the classical case, and hence any algorithm for the classical walk on the circle might have the potential to be performed quadratically faster with a quantum computer \cite{Aharonov2001}. \\

Studies have been made on the effects of controlled decoherence on the system, bridging the difference between the quantum and classical worlds. For a walk on the line, the quantum system is highly susceptible to decoherence, becoming essentially classical even if there is only very little noise in the coin space \cite{Kendon2003}. Despite this fragility, the powerful potential of QRW is well illustrated by the algorithmic speed-up which has been theoretically established. For example, Childs \textit{et al.} \cite{Childs2003} have shown that for a general graph traversal problem, the (continuous) QRW can be used in a way which is better than any possible classical algorithm. This is useful since many problems can be mapped onto such a graph traversal problem, including decision problems \cite{Childs2003}. \\

These traits all help define QRWs, yet there are many variations on the graphs being walked and how they are being walked. Some QRWs do not have all of these properties, so it is not possible to make a list of quantitative parameters to compare and pick a QRW which will be most suitable for quantum computing. Therefore, there is a continuous interaction between investigations into promising theoretical constructions and the physical constraints which define how a QRW will behave if experimentally realised. \\

\section{Optical quantum processing}

This project is aimed towards investigating a physical platform of quantum information processing using optical devices. Currently, using single photons as qubits has emerged as a leading approach to physically realise a quantum computer \cite{OBrien2007}. A significant advantage of single photons as qubits is that they are less prone to noise or decoherence than other systems under investigation. However, a significant disadvantage is that interfering photons with each other is all the more difficult than other quantum systems. Photons have several degrees of freedom such as polarization, frequency and spatial mode (i.e. the path of propagation). Each of these can be used to store information. If the information is stored in the polarisation of a photon, we say that the qubit is polarisation encoded. If the information is stored in the path taken by a photon, we say that the qubit is spatially encoded. \\

\begin{figure}[hbt]
  \begin{center}
    \resizebox{!}{30mm}{\includegraphics{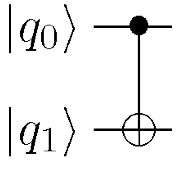}}
  \end{center}
  \caption{The Controlled-NOT gate acts on the target bit $q_1$ if and only if control bit $q_0 = 1$. (Reproduced from \cite{Chung2008}. ) }
  \label{CNOT}   % <-- IMPORTANT: \label is *after* \caption
\end{figure}

\begin{figure}[hbt]
  \begin{center}
    \resizebox{!}{40mm}{\includegraphics{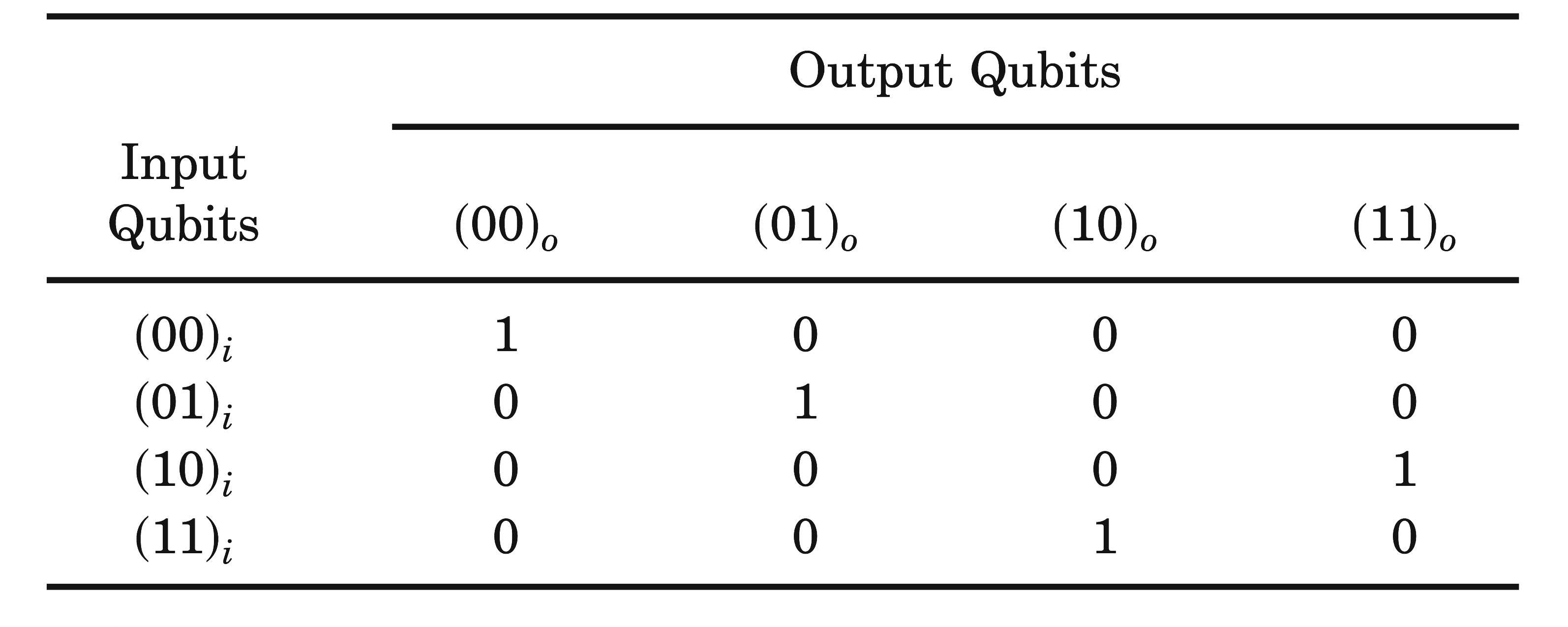}}
  \end{center}
  \caption{The Controlled-NOT gate truth table. (Reproduced from \cite{White2007}.)}
  \label{CNOTtruth}   % <-- IMPORTANT: \label is *after* \caption
\end{figure}

In quantum computing, qubits are manipulated by gates just like bits are in classical computing. Of course, qubit gates interact with qubits differently from how classical gates interact with classical bits. In classical circuit logic, all logic gates can be generated by various combinations of the NAND gate. In quantum computing, there are single-qubit gates that modify individual qubits. Also, there is the two-qubit CNOT (controlled-NOT) gate as shown in Fig.~\ref{CNOT}, which is a fundamental qubit gate. The truth table for the CNOT gate is displayed in Fig.~\ref{CNOTtruth}. Any other multi-qubit gate can be generated by combining CNOT gates and single-qubit gates \cite{Barenco1995}. Therefore, the CNOT gate is considered a vital part of implementing quantum computing. We also look at it here because a CNOT gate is an example of an important quantum device that can be created in an integrated optical chip.\\

\section{Bulk optics devices}

Initial investigations of optical quantum information were performed with bulk optics equipment. Therefore, integrated photonics devices often emulate bulk optics devices, as we will see later. However, to understand this relationship, we here introduce the bulk optics device which is emulated in the project. In bulk optics, a 50-50 beam-splitter (BS) is a partially reflecting mirror that allows half the intensity of incident classical light to pass through the mirror, and reflects another half of the light in another direction.  
\\

\begin{figure}[hbt]
  \begin{center}
    \resizebox{!}{60mm}{\includegraphics{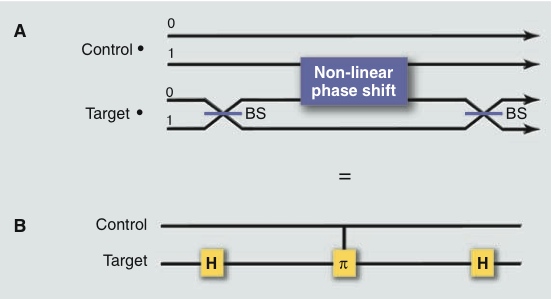}}
  \end{center}
  \caption{The Controlled-NOT gate could possibly be optically implemented as in A. Equivalently we write this form of the CNOT gate as in B, in quantum circuit notation, where qubits always move forward in time from left to right. (Reproduced from \cite{OBrien2007}.)}
  \label{CNOTobrien}   % <-- IMPORTANT: \label is *after* \caption
\end{figure}

Using polarisation encoded qubits, single qubit logic gates are relatively easy to implement through birefringent wave plates, and it is also possible to convert between polarization and path encoding using a polarizing beam splitter \cite{OBrien2007}. However, creating an optical CNOT gate represents more of a challenge. This is theoretically possible by sending the control photon and target photon together through a nonlinear phase shifting medium. Note that the target qubit is operated on by Hadamard gates before and after the phase shifting medium, as in Fig.~\ref{CNOTobrien}. The Hadamard gates can each be implemented as a 50-50 BS \cite{OBrien2007}. 
 
However, the optical CNOT conception described above and in Fig.~\ref{CNOTobrien} requires a material with strong single photon nonlinearity, but no such material exists. In integrated photonics, nonlinear optics is currently impossible. Some solutions have been developed. Knill, LaFlamme and Milburn \cite{Knill2001} have devised a robust scheme of linear optics quantum information processing, consisting of BS, phase shifters, memory and sources and detectors of single photons. In this scheme, the nonlinearity required to perform the CNOT operation is induced by the measurement \cite{Knill2001, Scheel2003}. 
A probabilistic CNOT gate operates correctly as a CNOT gate with probability less than $100$\%. However, it is known which events succeed and which fail, so it is possible to only measure successful operations of the CNOT gate. With various repetitions and modifications such as quantum teleportation or using auxiliary photon measurements, the probability can approach $1$ but huge numbers of entangled states must be consumed for each CNOT operation to get a reliable outcome \cite{OBrien2007}.  
 
Nevertheless, it has been demonstrated that scalable optical quantum computing is feasible using only single photon detectors, simple (i.e. linear) optical circuits made up of BSs and single-photon sources \cite{Knill2001}. Scalability is very important in order for an implementation of quantum computing to be cost- and space-efficient enough to fully begin to realise the potential of quantum computing. However, due to the probabilistic nature of linear optical quantum computing gates, this scheme will require a large amount of resources to scale up. Various architectures are being investigated to balance out the resources required in a way that maintains the quantum computing advantage \cite{OBrien2007}.  \\

In one such scheme, a single informational qubit can be encoded into several physical qubits. In other words, the vulnerability of the information is reduced by encoding it redundantly. Thus after measurement the original qubit is still recoverable \cite{Knill2001, OBrien2007}. These qubits are entangled and therefore need entangling gates to be formed. By using more than $10 000$ entangled photon pairs one scheme managed only to have $95\%$ deterministic CNOT gate \cite{OBrien2007}. Clearly such resource overhead can be considered impractical by today's standards. \\

Optical quantum computing also relies on photon sources that can generate single photons to very stringent specifications. Firstly, to interfere quantum mechanically, two photons must be mutually indistinguishable in every degree of freedom. Therefore, to have the necessary control over the qubits, independent sources must be able to generate single photons which are indistinguishable.  This has been achieved with trapped atoms and ions. Colour centres in diamond are a likely candidate for single photon sources, combining the benefits of solid-state and atom-like energy properties \cite{OBrien2007}. However, the principal source of single photons remains the single-photon downconversion (SPDC) sources. \\

As for detectors, current detectors have an efficiency of only approximately $70\%$ and only very specialised detectors can distinguish between one, two and three photons in a single mode \cite{OBrien2007}. 

Another challenge is the downconversion sources, nonlinear optical crystals which, when suitably driven, produce photon pairs. Barbieri \textit{et al.} \cite{Barbieri2009} state that the primary problem with optical quantum gates is the downconversion sources of single photons which are not ideal and hence cause low fidelity either due to higher-order photon terms or noise due to reduced brightness in photon sources. \\

However, work is also being done to improve detector technology and superconductor-based devices may be the solution. To make optical quantum computing scalable, both sources and detectors need to reach a sufficient level of development, detecting and generating an arbitrary number of degenerate photons. The circuits for the photons are assumed to necessarily consist of low-loss microscopic optical waveguides, and integrated optics is the primary hope. Still, the final integration of sources, detectors and integrated waveguide circuits remains to be achieved. It might also be possible to store quantum information in the single-photon sources, such as in the spin of the colour centres in diamond \cite{OBrien2007}. If possible, the interaction of stationary and traveling qubits in the form of atoms and photons could have great enabling implications for quantum computing \cite{Ralph2006}. \\

\section{Integrated Quantum Photonics}

For quantum computing to be a useful technology, it must have fidelity and fault tolerance. In other words, the instruments must produce less errors and noise, and ideally be robust to environmental sources of noise. To satisfy these conditions, an integrated quantum photonics approach may prove very attractive. We can say that a device is monolithic if it is made from a single unit. A monolithic integrated circuit which will possess no bulk parts, should minimise any noise in the system.  In integrated waveguides, it is possible to perform Hadamard operations not by beamsplitters, but through directional couplers. \\

Couplers are devices that take advantage of the fact that when two sufficiently close waveguides are built in parallel in an integrated circuit, any light passing through one waveguide has a certain probability of tunnelling across from that waveguide to the other. The probability depends on the separation of the waveguides, as well as distance the light travels in parallel with the alternate waveguide and the difference in refractive index, or refractive contrast between the waveguides and the surrounding medium. With classical light, this translates into a proportional exchange of intensity of light from one waveguide to the other. When photons couple from one waveguide to the other, they do so probabilistically. In other words, the photon has a certain probability of existing in one waveguide, and as it couples into the next, the probability of existing in the first waveguide decreases. Simultaneously, the probability increases that the photon exists in the other waveguide. It's important to note that given an indefinite length of propagation along two coupling waveguides, the intensity of light never evens out between the two waveguides, but instead continues to couple back and forth, alternating the position of the maximum intensity laterally with respect to the light's direction of propagation. Therefore, coupling is a unitary and reversible process.  \\

Again, if the photons passing through the circuit are non-degenerate, i.e. distinguishable in any degree of freedom, the photons will interact with the waveguide and each other as classical particles. However, if they are degenerate, then the photons can interfere quantum mechanically. Exploiting these interferences is the basis of building logic gates and other quantum information devices in integrated quantum photonic circuitry \cite{Marshall2009}. \\

\section{Laser Direct-Write Fabrication Technique}

Using the laser direct-write technique to create integrated waveguide circuits in fused silica (SiO$_2$) glass, Marshall \textit{et al.} \cite{Marshall2009} have investigated the fabrication and measurement of photonic quantum circuits. This project builds on that work, using fused silica glass because it is an excellent low loss optical material and is robust. There is a reason why this technique has been chosen. In order to create integrated waveguide circuits, currently there are two fabrication techniques available: Lithography and the laser direct-write technique. \\

Lithography is an iterative process where chemical vapour deposition, UV irradiation over circuit masks and etching are separate and necessary steps. The lithographic technique for producing waveguides in glass is similar to the techniques used to produce semiconductor transistor chips in the conventional computer industry. A great advantage of lithography is that once the fabrication equipment and correct circuit design are available, a large volume of chips with extensive, complex circuits can be efficiently fabricated, in a manner suitable for mass production. In addition to this physical advantage, since lithography builds on techniques established by the semiconductor industry, there is a lot of expertise behind this approach. However, lithography also has disadvantages. The fabrication equipment is expensive. The circuit design of a device can not be modified after it has been created. Thirdly, lithography creates patterns in layers, and as such the technique is intrinsically limited to two-dimensional circuit fabrication. \\

Theoretically, lithography could create blocky three-dimensional structures by coordinating superimposed two-dimensional layers. However, with the direct-write technique, waveguides can be produced which are smooth and contiguous and which form a path in any direction, in all three spatial dimensions. The laser direct-write technique is fairly slow compared to optimal lithography in terms of how much waveguide circuitry can be produced at a given rate. However, unlike lithography, where the design of the entire circuit must be complete before fabrication, the direct-write technique is versatile in that one can create the waveguide circuits in parts, iteratively testing, investigating and changing different designs or parts thereof in the same monolith. Another advantage of the direct-write technique is that the necessary production equipment is inexpensive compared to lithographic fabrication equipment \cite{Marshall2009}. For the purposes of this project, the three-dimensional writing capability is the crucial advantage of this technique. \\

How, then, does the laser direct-write technique work? Monolithic integrated circuitry can be directly written in paths which are programmed by using computer-aided design software. The laser direct-write technique used in this study uses short, femtosecond long pulses to modify the chemical structure of glass in a focus point. The modification locally changes the refractive index in the glass permanently, but without damaging the glass in a way that prevents the propagation of light. This requires balancing the power level of the laser between what gives no effect and the power level where the glass gets damaged. \\

\begin{figure}[hbt]
  \begin{center}
    \resizebox{!}{75mm}{\includegraphics{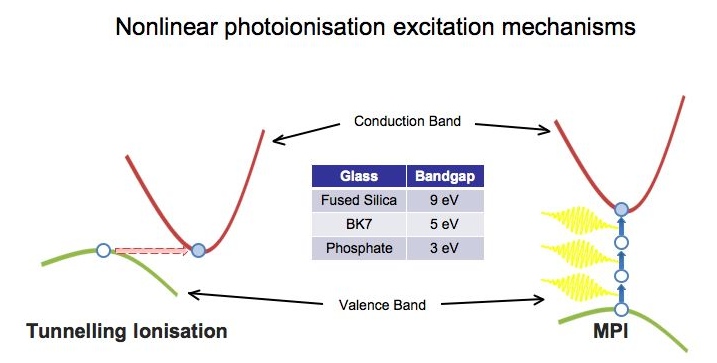}}
  \end{center}
  \caption{Two processes are involved in the refractive index modification. Incident photons cause tunneling ionisation and multi-photon ionisation, which in turn causes chemical and hence structural changes. (Provided by G. D. Marshall.)}
  \label{NPEM}   % <-- IMPORTANT: \label is *after* \caption
\end{figure}

The femtosecond laser direct-write technique can create a the refractive index modification in glass which is only located at the laser focus. The interaction between the laser and the glass is highly non-linear with laser intensity, and thus enables this precision. The laser-glass interaction initiates various photoionisation processes as shown in Fig.~\ref{NPEM}, such as multi-photon ionisation (MPI) by absorption, or tunnelling ionisation \cite{Siiman2009}. All these processes require a very high intensity which can only exist at the focus of the laser. For example, a bound electron in fused silica glass can be ionised by MPI when approximately six \unit{800}{\nano\meter} photons --- each with an energy of 1.55 eV --- are simultaneously absorbed to bridge the 9 eV band gap of this particular material. Thus liberated, a free electron can modify local bond structures. In this case, the bond structures are modified to densify the material and thus locally increase the refractive index.

As the refractive index is modified in the focus of the femto-second laser pulse, the glass can be moved in three dimensions around the focus to create waveguides inside the glass. For effective waveguides, several passes at the appropriate power level are required, where the glass is repeatedly moved in the same way relative to the focus point. This builds up the refractive contrast. The refractive index modification initially forms a Gaussian distribution about the focus point. As the refractive contrast builds up with repeated passes, the refractive index profile becomes more like a step-function. However, the Gaussian distribution of the direct-write effect about the focus volume means that the waveguide is not as distinct as an optical fiber. This represents a problem as we shall see in the next chapter, because there are some differences on the modelled systems' behaviour based on assumptions about the refractive profile of the waveguide. Previous work has led to refractive contrast profile models that are based on either a Gaussian distribution, or a pure step function. \\

The direct-write technique and equipment used in this project have enabled the creation of waveguides with specific properties. The waveguides that are created using the femtosecond laser direct write technique are equivalent to ones that can be accurately modelled in RSoft using a Gaussian profile, a width of \unit{2.972}{\micro\meter} and a refractive contrast of 0.00455. 

\section{Summary}

In this chapter we have reviewed the topics of quantum random walks, optical quantum processing, integrated photonics and the femtosecond laser direct-write technique. In the current project, all the areas discussed in the preceding sections are brought together. The intersection of these topics has already been partly investigated by others, such as Bromberg \textit{et al.}. In the following chapter, we begin a more detailed theoretical description of the light propagation in integrated photonics. \\

%**********************************************************************
%
%												Chapter 3 - Theory and simulations
%
%**********************************************************************

\chapter{Theory and simulations}

\section{Overview}

This chapter develops the the theory which underlies the practical realisation of this project. In the previous works by Bromberg \textit{et al.} \cite{Bromberg2009}, Hillery \textit{et al.} \cite{Hillery2003}, and Wang \textit{et al.} \cite{Wang2009}, several considerations were made that are relevant to this project, and these are explained below. It is also explained how the findings of work by others can be adapted to the particulars demands of this project. Specifically, this chapter present underlying theory, exploratory simulations and the rationale behind certain design decisions. 

\section{Theory}

This section explains how the simple QRW on the line can be translated into light propagation between waveguides, and presents details from the work by Bromberg {\it et al.} which inform this project. 

\subsection{Waveguides as beamsplitters}

\begin{figure}[hbt]
  \begin{center}
    \resizebox{!}{100mm}{\includegraphics{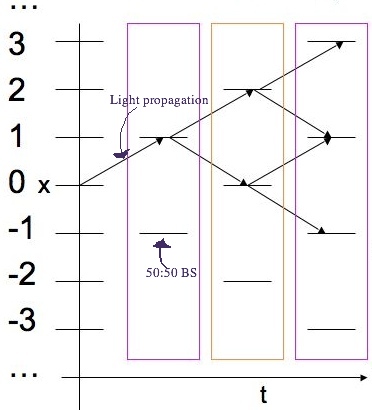}}
  \end{center}
  \caption{The mapping of the coined QRW on the line onto light propagating through a network of 50:50 beamsplitters. As the light propagates through the BS network, the vertical axis maps to the position x for the walker on the line, but the horizontal axis maps to time for the walker on the line. See text for a more full description.}
  \label{lineBSnet}   % <-- IMPORTANT: \label is *after* \caption
\end{figure}

This project is centred on a three-dimensional tubular device designed to realise continuous QRWs. Before describing the three-dimensional QRW, we first develop a one-dimensional description. Hillery \textit{et al.} \cite{Hillery2003} explore how the QRW on a circle can be expressed by circle of beamsplitters (BS), and this can be translated into a network of BS as shown in Fig.~\ref{lineBSnet}. They have studied transition rules and operators that express the quantum mechanics of a single photon traversing such a circle, and this work can provide a foundation for understanding how the  single photons will act in a well formed tubular waveguide array.

\begin{figure}[hbt]
  \begin{center}
    \resizebox{!}{50mm}{\includegraphics{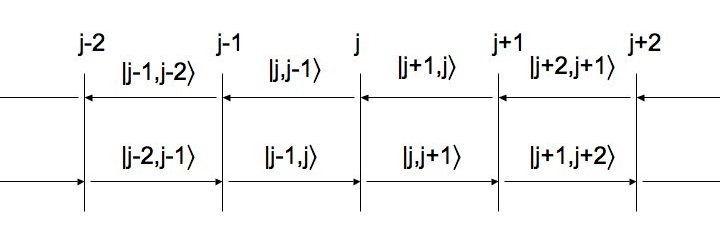}}
  \end{center}
  \caption{Visualising the QRW as a walk on a line of beamsplitters, as each photon transitions, its state is denoted by the vertex or BS it is going from and the vertex it is going to.}
  \label{BSline}   % <-- IMPORTANT: \label is *after* \caption
\end{figure}

The random walk on the line can be mapped onto a quincunx patterned network of beamsplitters, as shown in Fig.~\ref{lineBSnet}. The $z$-direction of light propagation in the BS network represents the dimension of time $t$ in the simple QRW on the line. When light hits a partially transmitting mirror (i.e. a beamsplitter), it may either transmit or reflect. Considering a quantum walk of photons transitioning between BS in a line, the state of each photon is denoted $| j,k \rangle$, where $j$ is the vertex which the photon is going from, $k$ is the vertex which the photon is going to, and $k=j \pm 1$ \cite{Hillery2003}. Now, if a photon in state $| j-1,j \rangle$ hits a BS, it either reflects and becomes $| j,j-1 \rangle$ or it transmits and becomes $| j,j+1 \rangle$. This is shown clearly in Fig.~\ref{BSline}.

 Furthermore, we can express the transition rules as follows \cite{Hillery2003}, with reference to Fig.~\ref{BSline}, that if the photon is incident on a BS $j$ from the left, then \\
 
 \begin{equation}
 | j-1,j \rangle \rightarrow t| j,j+1 \rangle+r| j,j-1 \rangle,
 \label{goright}
 \end{equation}
  
  and when the photon is incident on BS $j$ from the right,
  
 \begin{equation}
 | j,j+1 \rangle \rightarrow t^*| j,j-1 \rangle + r^*| j,j+1 \rangle.
 \label{goleft}
 \end{equation} 

The complex numbers $r$ and $t$ give the reflection and transmission coefficients respectively for each BS, such that $|r|^2+|t|^2=1$ Each BS can be hit by a photon from two sides, so the coefficients are present twice in the following block matrix, which represents each BS, \\
\begin{equation}
B_{BS}=  \begin{bmatrix} -r* & t \\ t* & r \end{bmatrix} .
\label{block}
\end{equation}

Hence, each block matrix $B_{BS}$ enacts the transition rule on an incident photon from either side.

This replaces the coin in the simple QRW on the line. Each column of BSs, shown as coloured rectangles in Fig.~\ref{lineBSnet}, can be represented as a unitary operator where the number of the column from the starting point determines if the unitary operator is odd or even. The unitary operators representing the columns thus take the following form, odd and even respectively: \\

$\hat{U}_{odd}=  \begin{bmatrix} B_{BS} & 0 & 0 & 0 \\ 0 & B_{BS} & 0 & 0 \\ 0 & 0 & B_{BS} & 0 \\ 0 & 0 & 0 & B_{BS} \end{bmatrix} $

and

$\hat{U}_{even}=  \begin{bmatrix} 1 & 0 & 0 & 0 & 0 \\ 0 & B_{BS} & 0 & 0 & 0 \\ 0 & 0 & B_{BS} & 0 & 0 \\ 0 & 0 & 0 & B_{BS} & 0 \\  0 & 0 & 0 & 0 & 1 \end{bmatrix} $. \\

\begin{figure}[hbt]
  \begin{center}
    \resizebox{!}{70mm}{\includegraphics{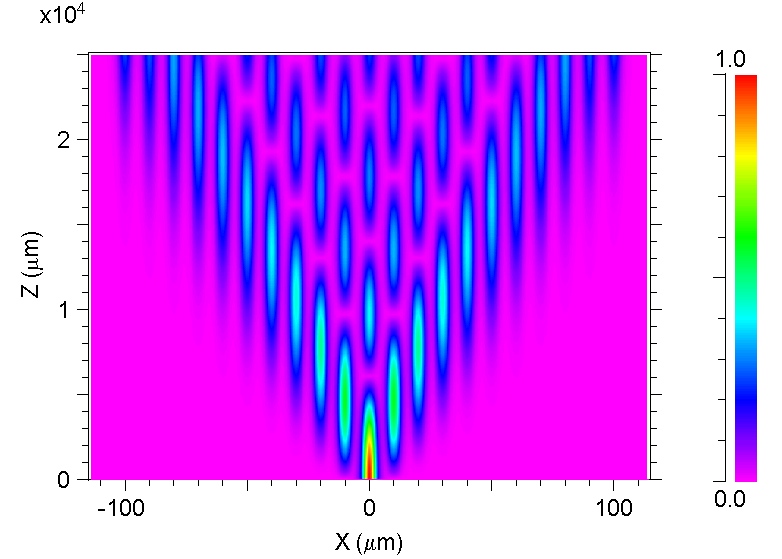}}
  \end{center}
  \caption{Light propagation through a planar waveguide array in the RSoft light simulator. The light is injected in the central waveguide at $x = 0$ and couples laterally in the $x$-direction while light propagates in the $z$-direction.}
  \label{RSoftplanarprop}   % <-- IMPORTANT: \label is *after* \caption
\end{figure}

\begin{figure}[hbt]
  \begin{center}
    \resizebox{!}{50mm}{\includegraphics{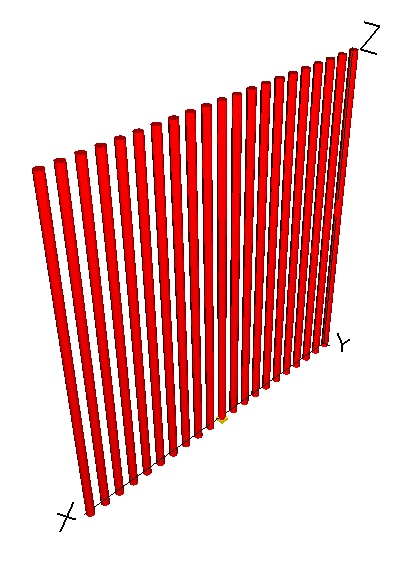}}
  \end{center}
  \caption{The three-dimensional display of a planar waveguide array RSoft light simulator.}
  \label{planararray3d}   % <-- IMPORTANT: \label is *after* \caption
\end{figure}

The network then acts as a succession of these odd and even unitary operators corresponding to the columns of BS in Fig.~\ref{lineBSnet}. Furthermore, when comparing the BS network and the propagation of light in a planar waveguide array, it is clear that this is the same pattern of light propagation, as shown in Fig.~\ref{RSoftplanarprop} and Fig.~\ref{planararray3d}. Consequently, a planar waveguide array can be equivalent to a BS network. This is explained by the fact that regular uniform parallel waveguides have a constant coupling rate, and hence after a certain distance, the light will have coupled $50 \%$ from one waveguide to the next. In other words, the light is coupled at regular distances of propagation, just as in the BS network in Fig.~\ref{lineBSnet}. Therefore, the quantum walk on the line is equivalent to the continuous QRW in the planar waveguide array. 

The equivalence of beamsplitters and waveguide couplers is important to enable our translation of the QRW into the waveguide picture. the equivalences mentioned above. When two parallel waveguides are close enough over a long enough distance, all the light will eventually couple across, and if the strucure continues, the light will couple back completely. After a distance where half the classical light has coupled across, if instead of classical light a single photon is traveling, the photon has a 50 \% chance of being in either waveguide. This is equivalent to a photon hitting a 50-50 BS and having a 50 \% chance of continuing its propagation on either side of the BS. In the quantum case, the quantum state remains coherent and the amplitude of the light splits into two smaller, equal peaks. \\

\begin{figure}[hbt]
  \begin{center}
    \resizebox{!}{80mm}{\includegraphics{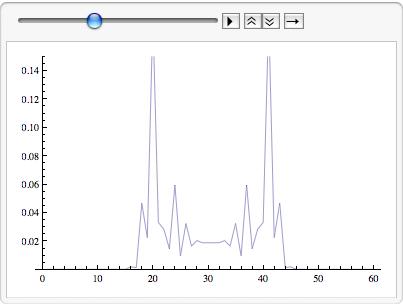}}
  \end{center}
  \caption{The QRW as it appears in a scattering through beamsplitters in a line. The underlying Mathematica code is based on the BS network from Hillery {\it et al.} \cite{Hillery2003} as described in this section. The resulting probability distribution is familiar, as in Fig.~\ref{Hypercube distribution}. (Code provided by J. Matthews) }
  \label{ScatteringQRWplay}   % <-- IMPORTANT: \label is *after* \caption
\end{figure}

Using the scattering QRW described by Hillery {\it et al.}, it is possible to produce a typical symmetrical QRW propagation, once again underlining the analogy between the QRW on the line and the QRW in a BS network. This is seen in Fig.~\ref{ScatteringQRWplay}. 

\subsection{Background from Bromberg {\it et al.}}

\begin{figure}[hbt]
  \begin{center}
    \resizebox{!}{60mm}{\includegraphics{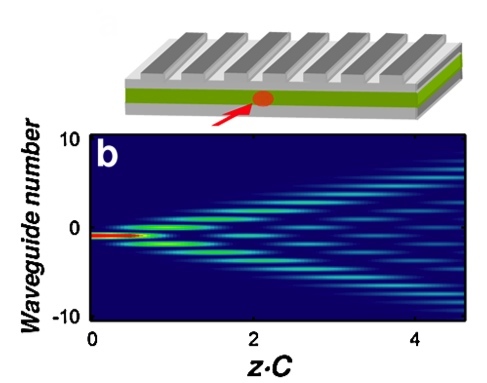}}
  \end{center}
  \caption{a.) The side view of a planar waveguide array. b.) The planar waveguide array, propagation of light with coupling. (Reproduced from \cite{Bromberg2009}.)}
  \label{brombergplanar}   % <-- IMPORTANT: \label is *after* \caption
\end{figure} 

What happens when photons propagate in a waveguide array? A waveguide array is a set of parallel waveguides that are close enough for light to couple between the waveguides, as shown in Fig.~\ref{brombergplanar}. The question was investigated by Bromberg \textit{et al.} \cite{Bromberg2009}, who simulated and experimentally observed the correlation of photon pairs propagating in a planar array of waveguides \cite{Bromberg2009}. In their work, a planar waveguide array had a pair of photons injected into either two adjacent waveguides, or next-to-adjacent waveguides. This was analysed in quantum mechanical simulations, but the experimental observations were made using bright classical light, to simulate the single photon experiments. This is possible because a single photon behaves like classical light when there is no interference. Furthermore, quantum simulations were done for the same 
setup but with path-entangled photon pairs. \\

\begin{figure}[hbt]
  \begin{center}
    \resizebox{!}{100mm}{\includegraphics{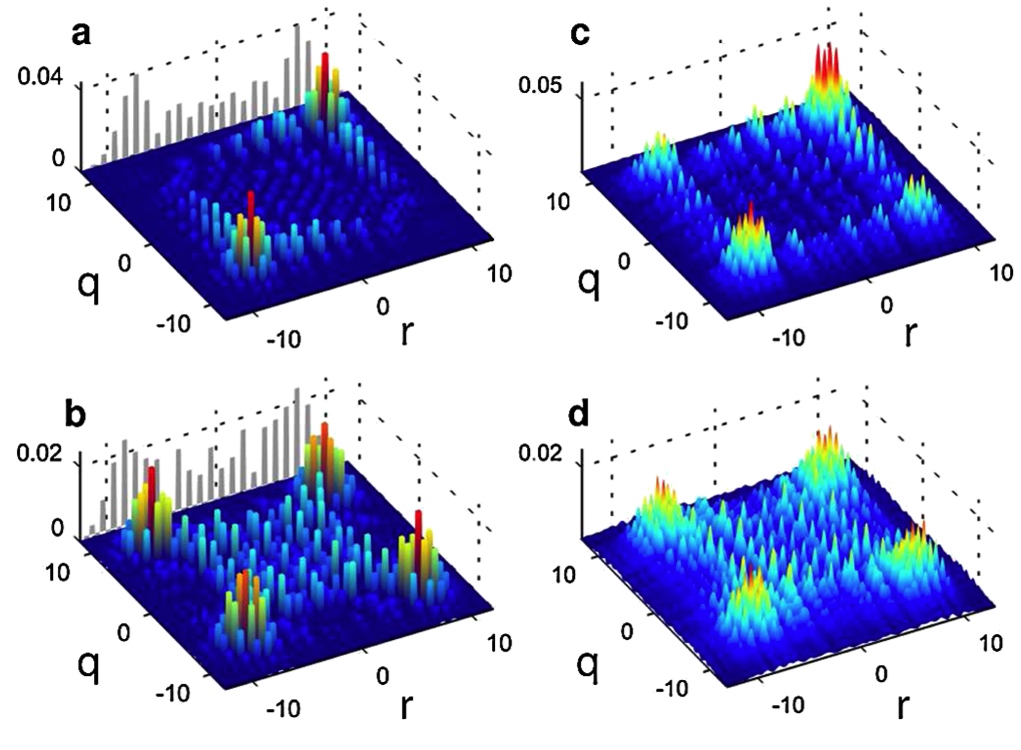}}
  \end{center}
  \caption{Correlation matrices: a.) Quantum simulation of two non-entangled photons injected into adjacent waveguides. b.) Quantum simulation of two non-entangled photons injected into next-to-adjacent waveguides. c.) Experiment with bright classical light injected into adjacent waveguides. d.) Experiment with bright classical light injected into next-to-adjacent waveguides. Each correlation matrix shows the probability that the detection of a photon at position $q$ at the end of the planar array should coincide with detection of the other photon at position $r$. In the classical case, the correlations reflect the equivalent coincidences of proportional intensities of light at the different exit points. (Reproduced from \cite{Bromberg2009}.)}
  \label{brombergnonentcorr}   % <-- IMPORTANT: \label is *after* \caption
\end{figure}

Both the simulations and measurements from Bromberg \textit{et al.} are focused on correlations, i.e. the degree of coincidence between different photons passing through the planar waveguide array and being detected upon exiting the array. Bromberg \textit{et al.} performed a classical experiment to simulate the injection of two separate single photons. The relationship between the correlation matrices of quantum simulations of non-entangled photon pairs and experiments with bright classical light is shown in Fig.~\ref{brombergnonentcorr}. 

\begin{figure}[hbt]
  \begin{center}
    \resizebox{!}{60mm}{\includegraphics{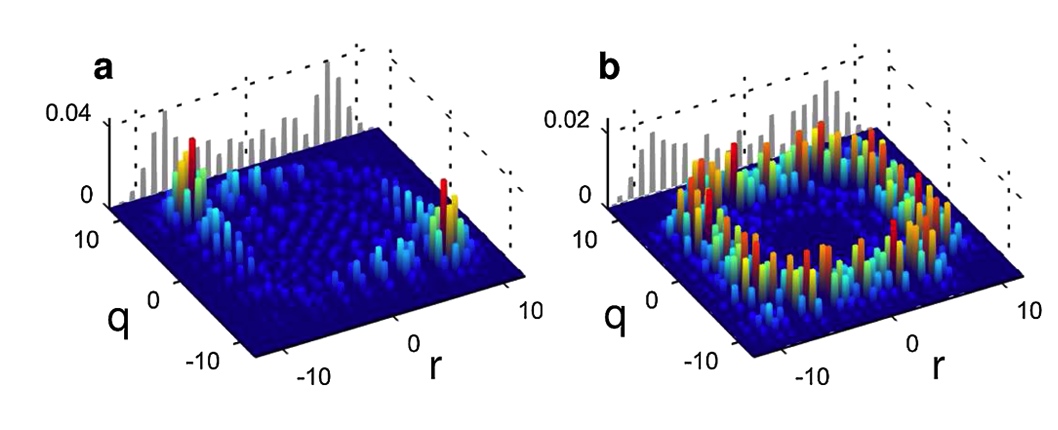}}
  \end{center}
  \caption{Correlation matrices: a.) Quantum simulation of two entangled photons injected into adjacent waveguides. b.) Quantum simulation of two entangled photons injected into next-to-adjacent waveguides. (Reproduced from \cite{Bromberg2009}.) }
  \label{brombergentcorr}   % <-- IMPORTANT: \label is *after* \caption
\end{figure}

However, for the entangled photon pairs, no classical experimental analogue was possible, and no single photon experiments were reported. The quantum simulations of the entangled photon pairs showed that when the photon pair is injected into either of two neighbouring waveguides, the two photons will always emerge at opposite sides of the array. On the other hand, if the entangled photon pair is injected into either of two next-nearest-neighbouring waveguides, one photon will emerge at a side, while the other always exits at the centre of the waveguide array, e.g. at $x =0$ inFig~\ref{RSoftplanarprop}. Note that the width of the array and the length of propagation is such that the photons never reach a boundary, which would reflect the coupling photon back towards the centre of the array. Indeed, boundary reflections are a fundamental limitation to studies of this kind in one-dimensional QRWs. The correlation matrices from the quantum simulations of entangled photons are shown in Fig.~\ref{brombergentcorr}. By comparison with Fig.~\ref{brombergnonentcorr}, Fig.~\ref{brombergentcorr} illustrates that the behaviour of non-entangled photon pairs is very similar to classical light. \\

The work of Bromberg, \textit{et al.}, shows some aspects of how waveguide arrays can be used to control and manipulate quantum states, which is fundamental to developing quantum computing \cite{Bromberg2009}. However, the extent of current knowledge in this are rudimentary, suggesting the need for further experimental investigations of light in a waveguide array. This project takes it a significant step further, by investigating a three-dimensional waveguide array. The topology of the QRW would be two-dimensional, rather than the one-dimensional walk investigated by Bromberg \textit{et al.}, but could not be studied using photon without the three-dimensional tubular array of waveguides. \\

\section{Simulations}

Before moving forward from the work done by Bromberg \textit{et al.} it has been important to establish a firm grasp on what has already been accomplished. By understanding how these results have come about, it is possible to appreciate the relationship between that study and this project. Furthermore, by seeing how this project can reproduce the Bromberg results directly, the way to expand on the previous work is clarified.  First the simple relationship between waveguide separation and the coupling rate is explored, without reference to previous work. Then some coupling-mode theory equations are introduced which give an alternative way of calculating the coupling process between parallel waveguides. Next, two correlation matrices from Bromberg {\it et al.} are reproduced by simulations with parameter values that the direct-write laser technique can realistically fabricate.  

\subsection{Side by side}

\begin{figure}[hbt]
  \begin{center}
    \resizebox{!}{60mm}{\includegraphics{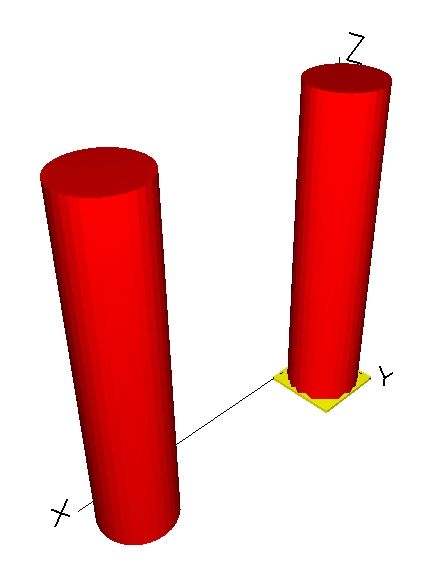}}
  \end{center}
  \caption{Two parallel waveguides in RSoft, as used to investigate the simplest case of waveguide coupling.}
  \label{sidebyside}   % <-- IMPORTANT: \label is *after* \caption
\end{figure}

\begin{figure}[hbt]
  \begin{center}
    \resizebox{!}{80mm}{\includegraphics{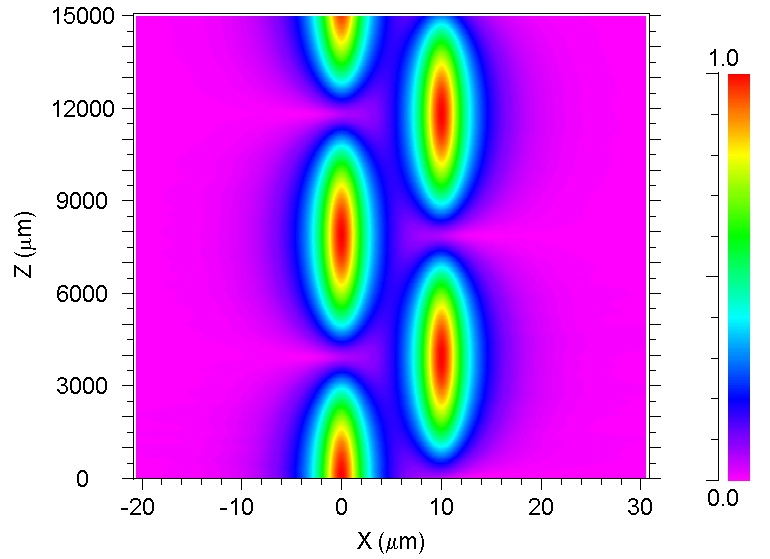}}
  \end{center}
  \caption{The propagation of light in two parallel waveguides in RSoft, as used to investigate the simplest case of waveguide coupling. Light with one unit of power is launched into the left waveguide (not shown) and this light couples between the waveguides in an oscillatorary manner. The 'physical' waveguides are not shown in this colour contour plot of power. }
  \label{sidebysideprop}   % <-- IMPORTANT: \label is *after* \caption
\end{figure}

We first consider two identical, parallel waveguides, as shown in Fig.~\ref{sidebyside}. Fig.~\ref{sidebysideprop} shows the propagation of light as it couples between two waveguides as computed by RSoft. Waveguide width and refractive contrast are two parameters which may not be easily varied using the direct-write technique. However, waveguide seperation can easily and reliably be modified in the designs in this project, with regards to coupling strength. This makes it important to be able to predict what impact a given waveguide separation has on coupling strength. \\ 

An empirical way to determine coupling coefficient between two waveguides can be found by using RSoft parameter scan, finding the length $L$ that the light travels before coupling completely across, and relating this to the separation $d$ between the two waveguides. We empirically found this relationship is expressed in the algebraic expression

\begin{equation}
L=87.988 e^{0.4005 d}. 
\label{sbseq}
\end{equation} 

\begin{figure}[hbt]
  \begin{center}
    \resizebox{!}{110mm}{\includegraphics{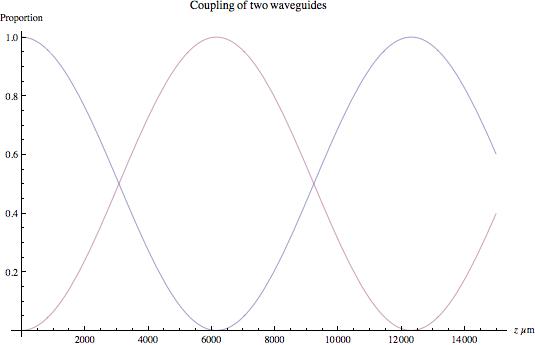}}
  \end{center}
  \caption{Using coupling mode theory (CMT) equations from Wang \textit{et al.} \cite{Wang2009} in Mathematica, to compare the two models. The distance between the waveguides is here \unit{10}{\micro \meter}.}
  \label{WangCMT2wgcoupling}   % <-- IMPORTANT: \label is *after* \caption
\end{figure}

\begin{figure}[hbt]
  \begin{center}
    \resizebox{!}{110mm}{\includegraphics{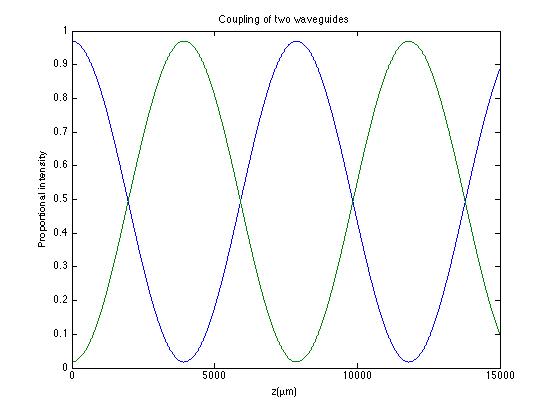}}
  \end{center}
  \caption{MatLab plot of RSoft results for a physical system ostensibly identical to the one modelled in Fig.~\ref{WangCMT2wgcoupling}. }
  \label{sidedbysidematlab}   % <-- IMPORTANT: \label is *after* \caption
\end{figure}

An empirical approach is not wholly satisfactory to gain deeper understanding of the physics involved and we subsequently look into Coupled-Mode Theory (CMT). Another way to explain this is to say that $L$ is the distance in z between peaks as the light couples back and forth between the two waveguides. However, although the sinusoidal shape is the same in Fig.~\ref{WangCMT2wgcoupling} and Fig.~\ref{sidedbysidematlab} we see some discrepancy between the two figures in the rate at which they oscillate. 

\subsection{Coupled-mode theory of parallel waveguides}

The equations used here to model the coupling of light between parallel waveguides in Mathematica are based on the coupling mode theory (CMT) as presented by Snyder \cite{Snyder1972}, Wang {\it et al.} \cite{Wang2009} and Yu {\it et al.} \cite{Yu2006}. Light propagating along a waveguide is expressed as a differential equation, and with the initial conditions one can solve for the light propagation in the waveguide array. For a system of $N$ (where $N > 3$) parallel fibers a planar array, the coupling is described by the system of equations 

\begin{equation}
\frac{\mathrm{d}A_1(z)}{\mathrm{d}z} + i \beta_1A_1(z)=-iA_2(z)C_{12} ,
\label{CMTDE1}
\end{equation}
\begin{equation}
\frac{\mathrm{d}A_m(z)}{\mathrm{d}z} + i \beta_mA_m(z)=-iA_{m-1}(z)C_{m(m-1)}-iA_{m+1}(z)C_{m(m+1)} ,
\label{CMTDEm}
\end{equation} 
\begin{equation}
\frac{\mathrm{d}A_N(z)}{\mathrm{d}z} + i \beta_NA_N(z)=-iA_{N-1}(z)C_{N(N-1)}-iA_{N+1}(z)C_{N(N+1)} ,
\label{CMTDEN}
\end{equation}

where $(m=2,3,...,N-1)$,

If $n_1$ and $n_2$ are the refractive indices of the fiber core and the cladding background, respectively, and the radius of each fiber core is taken as $r$, and $d$ the distance between the centres of adjacent fiber cores, then the following equations give the coupling coefficient $C$ and the mode propagation constant $\beta$ in terms of the three dimensionless parameters $U$, $V$, and $W$, assuming step index profiles:

\begin{equation}
\beta = \left[ \left( \frac{2 \pi n_1}{ \lambda{} } \right)^2 - \frac{U^2}{r^2} \right] \label{CMTbeta} \\
\end{equation}

\begin{equation}
C=\frac{\sqrt{\delta}U^2K_0\left[ W(d/r) \right]}{rV^3K_1^2(W)} \label{CMTC} \\
\end{equation}
where $K_0$ and $K_1$ represent modified Hankel's functions --- also known as the modified Bessel function of the second kind --- of the zeroth and first order, respectively. These expressions can be further expanded by the equations defining $\delta$, $U$, $V$ and $W$: \\
\begin{equation}
\delta = 1-\left( \frac{n_2}{n_1} \right) ^2 \label{CMTdelta} \\
\end{equation}
\\
\begin{equation}
V=\frac{2 \pi r n_1 \sqrt{\delta} }{\lambda} \label{CMTV} \\
\end{equation}
\\
\begin{equation}
V^2= U^2+W^2 \label{CMTVUWsq} \\
\end{equation}
\\
\begin{equation}
\frac{UJ_1(U)}{J_0(U)}=\frac{WK_1(W)}{K_0(W)} \label{CMTbessel} ,\\
\end{equation} \\

where $J_0$ and $J_1$ represent the zeroth and first order Bessel functions of the first kind. Then, given numerical values for all the physical parameters, Eq.~\eqref{CMTbessel} and Eq.~\eqref{CMTVUWsq}  can be solved to determine $U$, $V$, $W$ and then Eqs.~\eqref{CMTV} and Eq.~\eqref{CMTdelta} give $\delta$, and from there on $\beta$ can be determined with Eq.~\eqref{CMTbeta}. Finally, Eq.~\eqref{CMTC} gives coupling coefficient $C$. With the coupling coefficients and mode propagation constant thus determined, the differential equations for a particular system of parallel coupling waveguides can be solved for any particular point of propagation.  We wanted the light obey CMT so as to easily convert into quantum mechanics. \\

Thus equipped, it is possible to compare RSoft results with explicit CMT calculations in Mathematica. However, the discrepancy between the two is obvious, and the simulation work largely focusses on finding a set of parameters where the two simulation methods are in approximate agreement. Also, the simulation work initially sought the production of distinct recurrences, where after some propagation all the light would again exist in the waveguide where it has been injected. \\

\subsection{Correlation matrices emulating Bromberg {\it et al.}}

\begin{figure}[hbt]
  \begin{center}
    \resizebox{!}{60mm}{\includegraphics{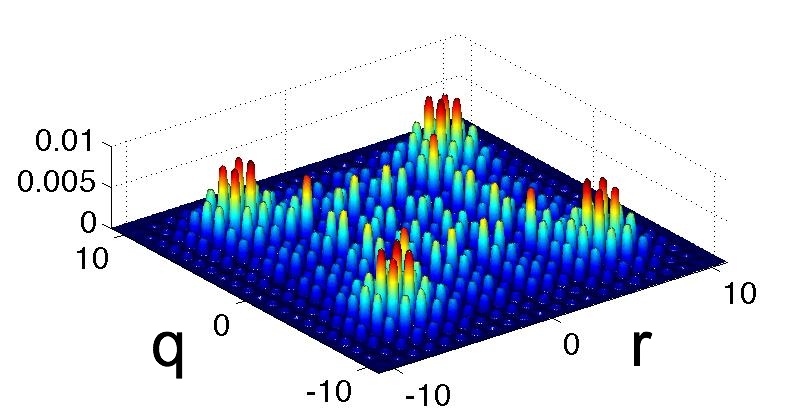}}
  \end{center}
  \caption{This correlation matrix results from injecting bright classical light into two next-to-adjacent waveguides and taking the average of the possible relative phases.}
  \label{bromulate2}   % <-- IMPORTANT: \label is *after* \caption
\end{figure}

Using the RSoft BeamPROP software, the correlation matrices from the classical experiments by Bromberg {\it et al.} \cite{Bromberg2009} are reproduced by phase averaging the classical light of one of the beams injected into the planar waveguide array. Light injected into two adjacent waveguides gives a correlation matrix as seen in Fig.~\ref{bromulate2}, which corresponds to Fig.~\ref{brombergnonentcorr} a.) and c.) from Bromberg {\it et al.} \cite{Bromberg2009}. Similarly, the emulation of Fig.~\ref{bromulate2} corresponds to Fig.~\ref{brombergnonentcorr} b.) and d.). By testing this we are better able to understand how to experimentally test three-dimensional waveguide tubes using bright light. 

\begin{figure}[hbt]
  \begin{center}
    \resizebox{!}{100mm}{\includegraphics{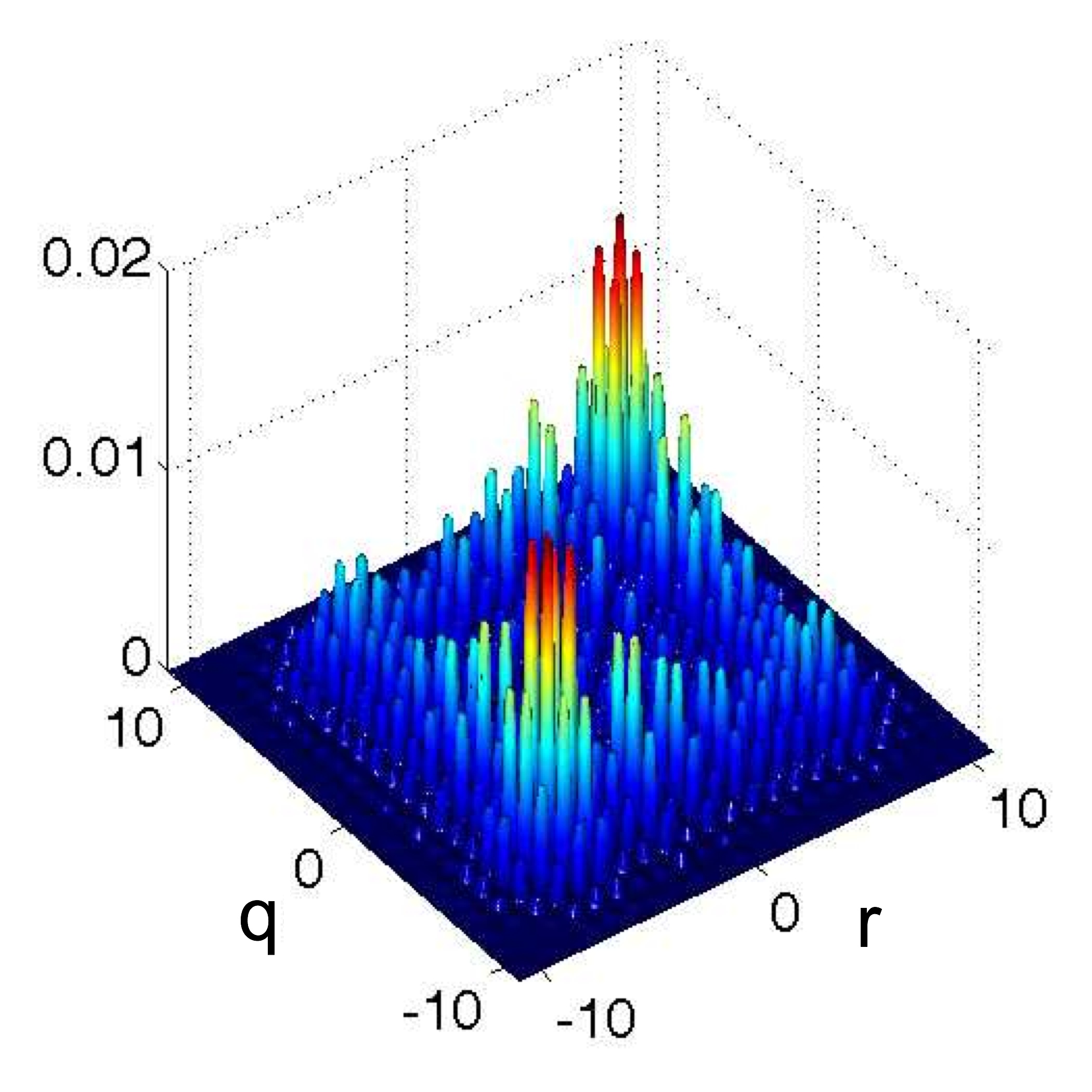}}
  \end{center}
  \caption{This correlation matrix results from injecting bright classical light into two adjacent waveguides and taking the average of the possible relative phases.}
  \label{bromulate1}   % <-- IMPORTANT: \label is *after* \caption
\end{figure} 

\section{Tubular waveguide array design}

Finally, the design decisions behind the fabricated devices are discussed. Three primary decisions have been made that determined the present outcome. First, the number of waveguides in the tube has been decided. Second, the radius of the tube was determined. Third, both a single-stage and two-stage fan-in section have been fabricated. The dilemmas and rationales behind these decisions are also presented. 
\\

\begin{figure}[hbt]
  \begin{center}
    \resizebox{!}{60mm}{\includegraphics{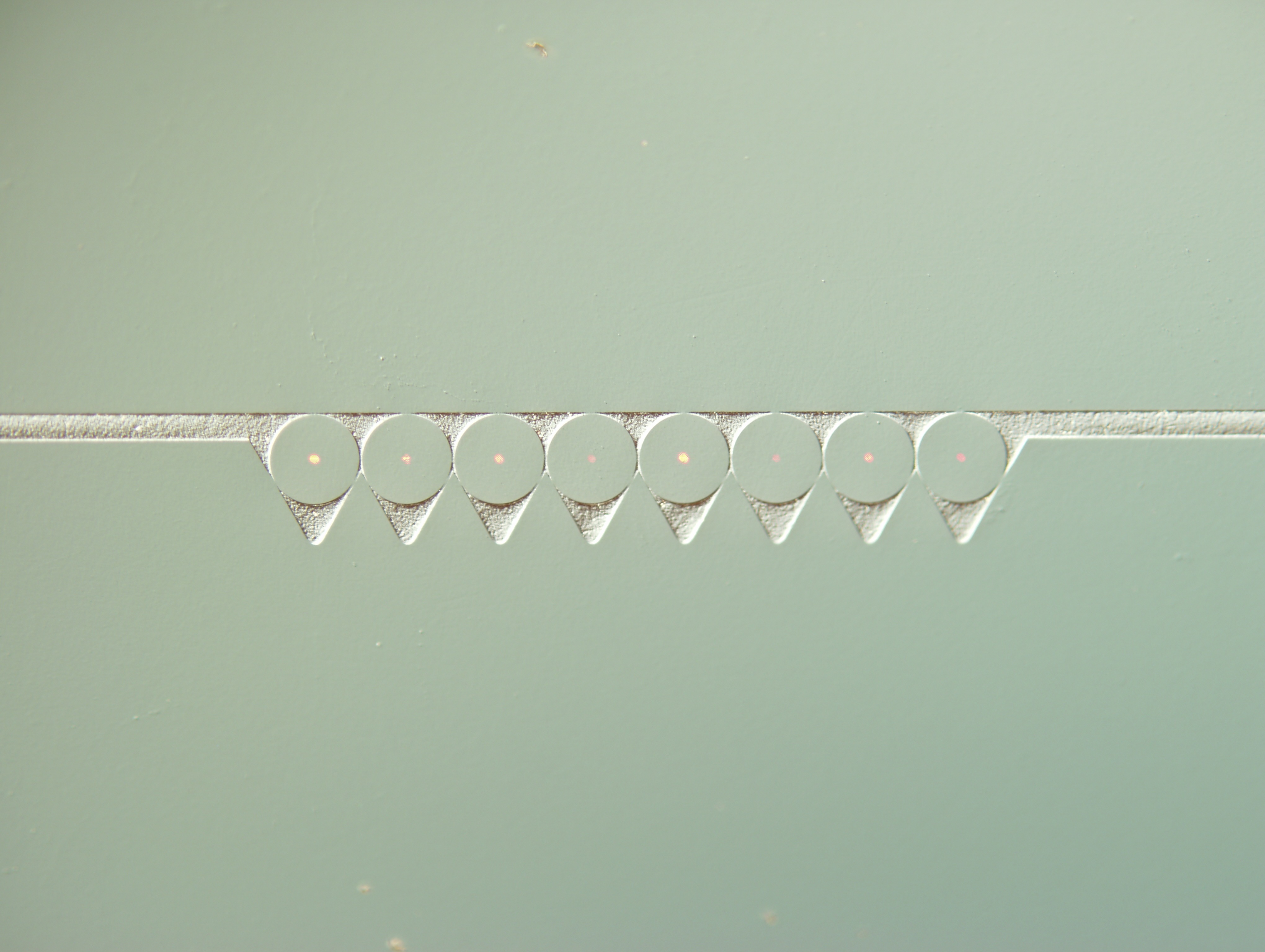}}
  \end{center}
  \caption{The V-groove array tightly binds optic fibers in parallel, equally spaced at precise \unit{127}{\micro\meter} intervals. This device can connect bulk optics with integrated photonics. (Provided by G. D. Marshall)}
  \label{vgroove}   % <-- IMPORTANT: \label is *after* \caption
\end{figure} 

Initially, the number of waveguides per tube was limited by the number of bulk optical sources which would be connected with the device via fiber optics. Because the V-groove array, shown in Fig.~\ref{vgroove} is the device which precisely spaces out optical fibers to connect bulk optics with the fan-in section of the integrated photonic device, the maximum number of light sources into the device is either eight or sixteen. With a view towards simplicity and producing recurrence, the limit of eight waveguides was initially accepted, and further reduced to six in an attempt to enable recurrence after a minimum propagation length, while still having enough waveguides to make the QRW an interesting prospect. Recurrence happens when the light has propagated far enough in the array that the coupling has come back to concentrate all the light intensity in the same waveguide where the light was initially injected.The reason why recurrence is interesting is that if we can demonstrate such a recurrence in the physical device, this will be a strong indication that the waveguides which make up the tube are indeed well-formed, parallel and have identical refractive profiles. \\

In order to produce a recurrence intensity distribution in the physical device, it's important that the recurrence occurs after a relatively short longitudinal propagation on the order of 20 mm. This is because the silica glass in which the device is direct-written has a maximum length of 30 mm, and the first one third will be occupied by the fan-in section. This length is required to have a fan-in section where the waveguides leading input light from the v-groove array arc gently, gradually enough towards the lateral (x-y plane) position of the tube, so that the transmission of light is not lossy. \\

To make the light propagation through a tubular waveguide array truly equivalent to a QRW on a circle, there can be no transitions between non-adjacent vertices on the circle. Ideally, there should be no non-nearest-neighbour coupling. However, during the simulations non-nearest-neighbour coupling (NNNC) has been found to occur to some extent in the tubular array for all relevant tube radii. With a view towards having a tubular array which can produce a recurrence within the extension of the integrated photonic chip, the length of propagation is also limited. Since the coupling strength decays exponentially with increasing waveguide separation, as implied by Eq.~\eqref{sbseq}, a larger tube radius could reduce NNNC, but would all the more counteract the possibility of recurrence within the tubular array at the available propagation length, which as it turned out is between \unit{20-22}{\milli\meter} depending on the fan-in section, but even if no fan-in section was required, the tube could not exceed \unit{30}{\milli\meter} in length. \\

\begin{figure}[hbt]
  \begin{center}
    \resizebox{!}{90mm}{\includegraphics{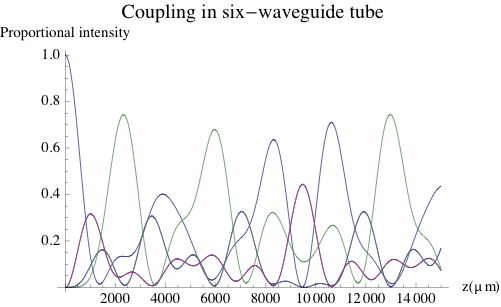}}
  \end{center}
  \caption{The coupled-mode theoretical calculations of the coupling of light in the six-waveguide tube with \unit{7}{\micro\meter} radius.}
  \label{WangCMT15000-7mum}   % <-- IMPORTANT: \label is *after* \caption
\end{figure} 

\begin{figure}[hbt]
  \begin{center}
    \resizebox{!}{110mm}{\includegraphics{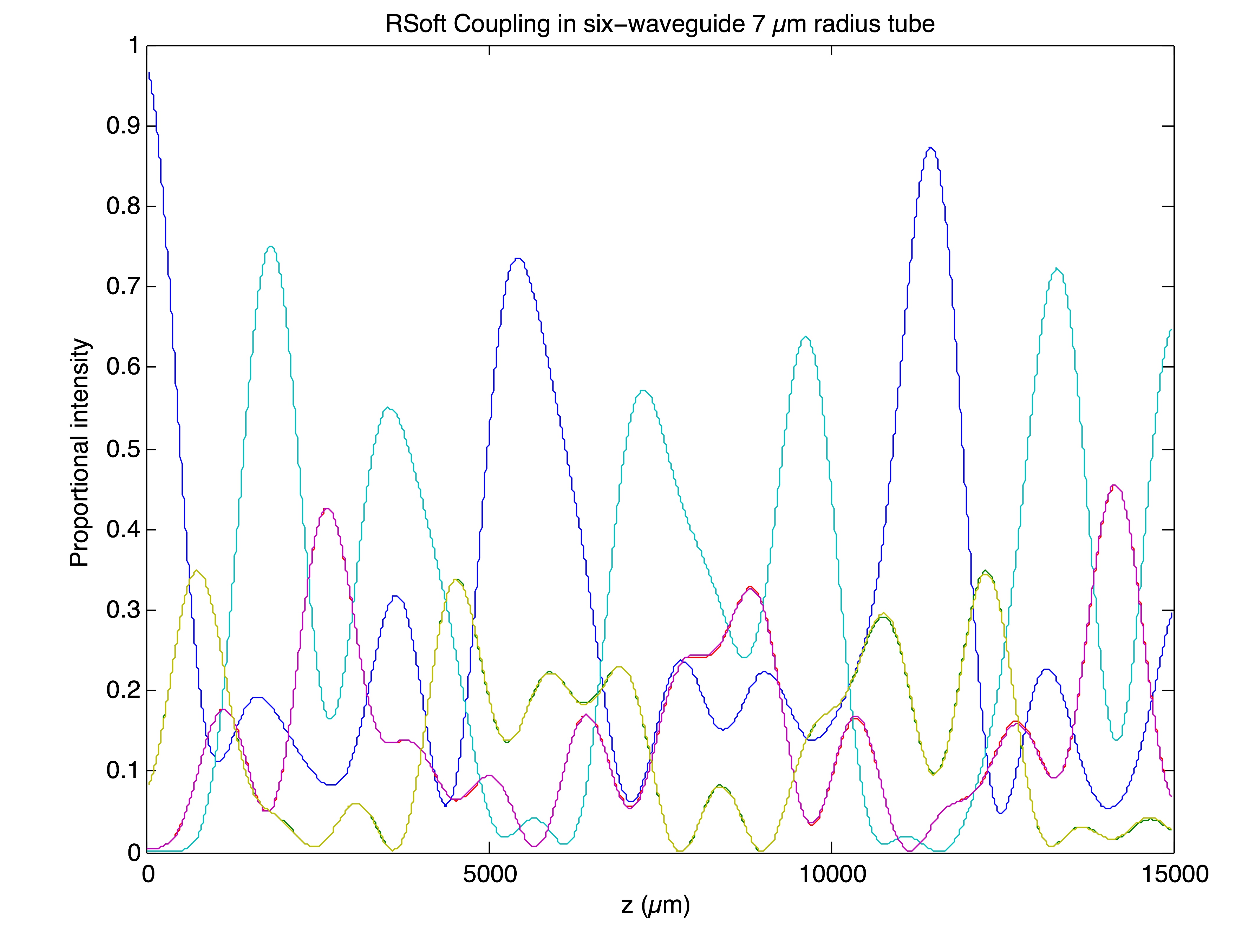}}
  \end{center}
  \caption{The coupling of light in the six-waveguide tube with \unit{7}{\micro\meter} radius, as calculated in RSoft BeamPROP.}
  \label{RSoft15000-7mum}   % <-- IMPORTANT: \label is *after* \caption
\end{figure}

So how was the \unit{7}{\micro\meter} tube radius determined? If the refractive contrast could be increased significantly, NNNC can be reduced. However, we have accepted the current development of the direct-write technique and base the design on established production capabilities. Thus the tube radius is decided by finding a radius value which gives both a relative agreement between the RSoft and Coupled-Mode Theory simulations and also with a recurrence-like maximum in the propagation length. However, there is still discrepancy, as can be seen in Fig.~\ref{WangCMT15000-7mum} and Fig.~\ref{RSoft15000-7mum}. Other radii have simply been found to produce more discouraging discrepancies, and ultimately a tube radius had to be decided so that fabrication could proceed.

It is important to note that while observing recurrence was initially an important objective, and although a complete recurrence could not be observed due to NNNC, there is another reason why recurrence measurement have not been performed. In order to measure the evolution of light in the tube, which would be necessary to observe the recurrence, destructive cut-back measurements would have to be made to the fabricated device. This is not an attractive prospect. \\

The fan-in sections have been designed to give  95 \% or greater coupling of light from the farthest input waveguides (1 and 6) to the tube section. The light travelling through these two waveguides exeriences the most lateral displacement, is subjected to the smallest bend radius, and hence the highest loss. The form of the curves is the raised sine, as described in the RSoft documentation. This form has been empirically demonstrated to be the lowest-loss type among many curves expressed in trigonometric or polynomial functions.  

\begin{figure}[hbt]
  \begin{center}
    \resizebox{!}{90mm}{\includegraphics{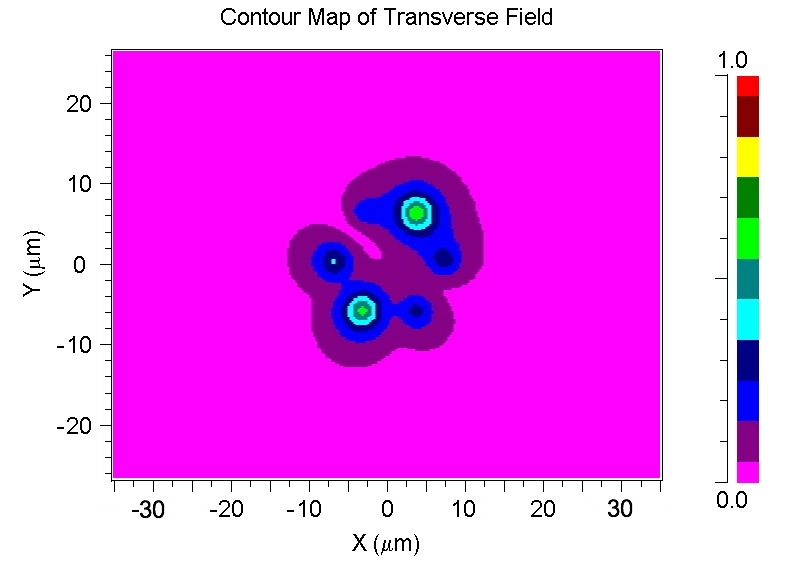}}
  \end{center}
  \caption{The coupling of light in the six-waveguide tube with \unit{7}{\micro\meter} radius, as calculated in RSoft BeamPROP. The slight rotational asymmetry is evident.}
  \label{asymmetric-1sfanin}   % <-- IMPORTANT: \label is *after* \caption
\end{figure}

\begin{figure}[hbt]
  \begin{center}
    \resizebox{!}{90mm}{\includegraphics{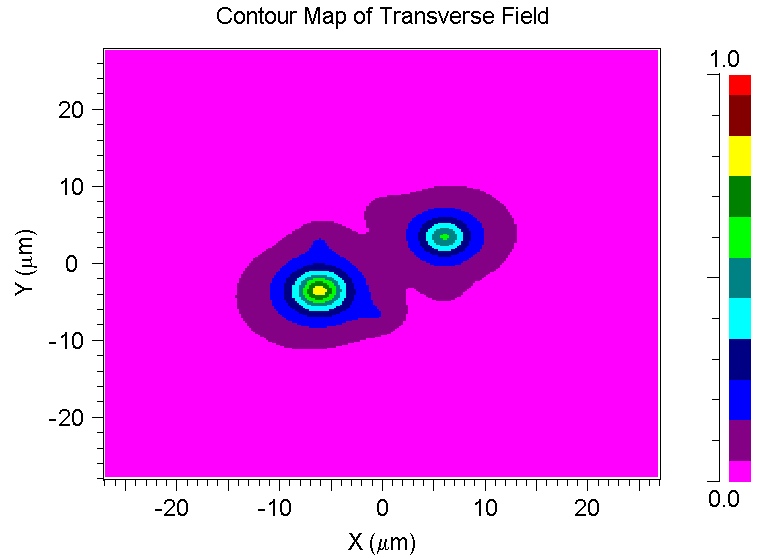}}
  \end{center}
  \caption{The coupling pattern of light in the tubular waveguide array with \unit{7}{\micro\meter} radius, as calculated in RSoft BeamPROP. Here, light is launched into one waveguide without a fan-in section, i.e. the light is symmetrically distributed as it enters the coupling region. }
  \label{symmetric-nofanin}   % <-- IMPORTANT: \label is *after* \caption
\end{figure}

The single-stage fan-in was found to have a slight rotational asymmetry in the way light propagates. Therefore, while a single-stage device has been fabricated, a second device was also created with a two-stage fan-in section. The first stage, like the only one in the single-stage fan-in case, is a fan-in from waveguides in a plane into a tube, but the tube is wider than in the coupling region. This second stage then contracts into the \unit{7}{\micro\meter} radius tube. Some interference and coupling might occur before the second stage has finished, but if so, it should at least be symmetric interference. The rotational asymmetry of the single-stage fan-in device has been simulated in RSoft, giving Fig.~\ref{asymmetric-1sfanin}. By comparison, if light is simply launched into one tube in RSoft, the coupling pattern is symmetric, as in Fig.~\ref{symmetric-nofanin}. 

%**********************************************************************
%
%												Chapter 4 - Methods and measurements
%
%**********************************************************************

\chapter{Methods and measurements}

The experimental stage of the project consists of fabricating the device that was designed in previous stages, and then measuring the intensity distribution of bright classical light after a known length of longitudinal propagation through the tubular array of waveguides. These measurements are accomplished with beam profilometry. Also, refractive contrast profilometry is used to determine the refractive contrast of the waveguides. However, the numerical calibration and analysis of these measurements are still pending. 

\section{Fabrication}

Using the laser direct-write technique, first one and then another tube design are fabricated in a slide of silica glass. The slide is mounted on a precise, mobile stage, and while the laser modifies the refractive contrast in the focal point, the stage moves in three dimensions according to the fabrication program being run. The laser is also blocked while the stage moves back, so that no unwanted refractive index change occurs. To build up refractive index changes to be significant enough to form functional waveguides, the laser passes over the same areas repeatedly. A single-stage tube created with eight passes has been ruined because of power fluctuations in the laser during production. However, a sixteen pass production of the single-stage tube has been successful. Also, a two-stage tube was successfully fabricated with eight passes. Both tubes have a radius of $7$ Äm, and the second stage of the two-stage device fans in to  \unit{7}{\micro\meter} from a second-stage radius of \unit{14}{\micro\meter}. The two devices differ in that their \unit{7}{\micro\meter} regions are approximately 20 mm and 22 mm long, respectively. After the waveguides are direct-written in the silica glass, the monolith/slide is polished so that the exit of the tubes can be see head on, and so that the input waveguides can be accessed with bulk optical fibers leading in light. \\

\begin{figure}[hbt]
  \begin{center}
    \resizebox{!}{110mm}{\includegraphics{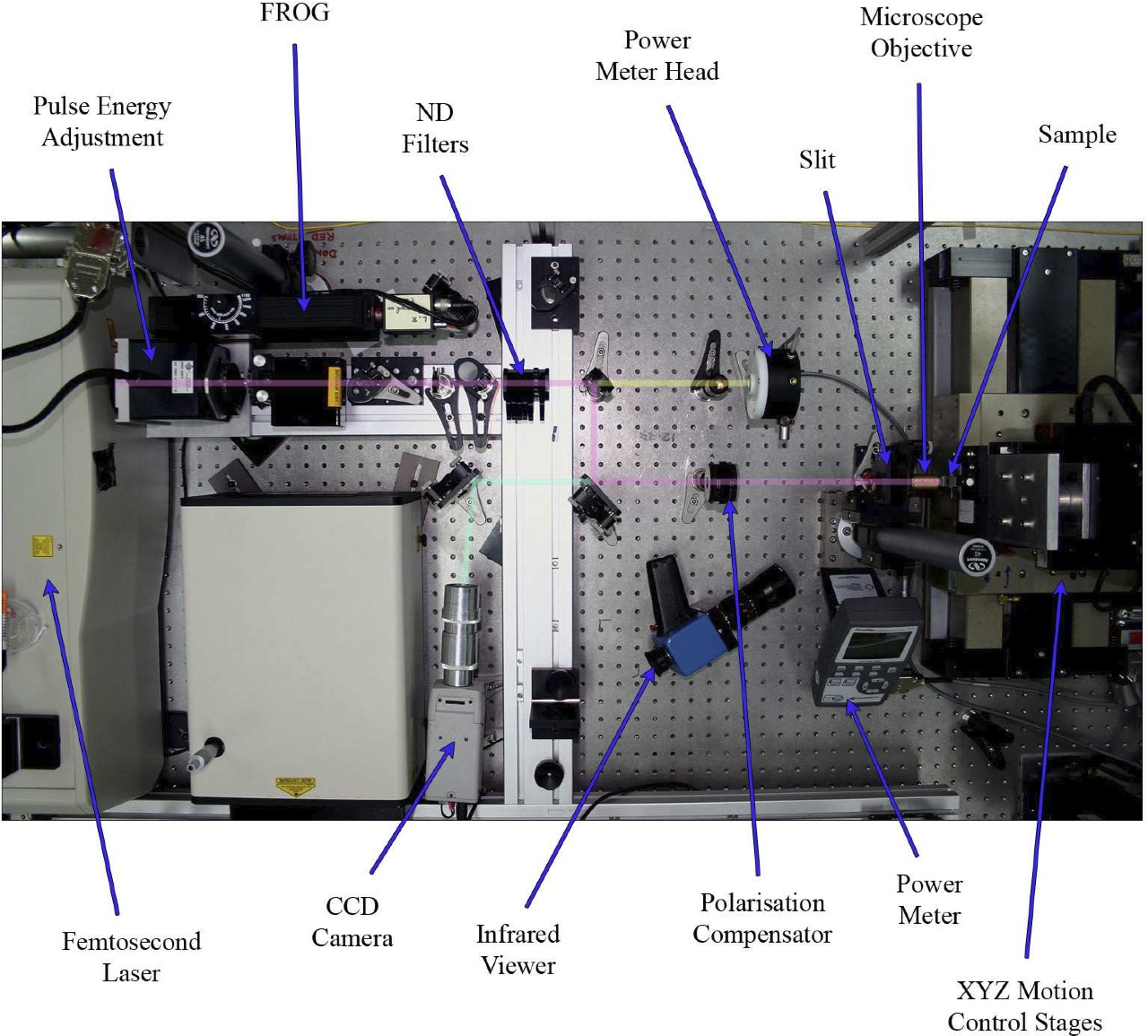}}
  \end{center}
  \caption{The setup of the optical bench during the direct-write fabrication procedure, seen from above. (Provided by M. Ams.)}
  \label{writepic}   % <-- IMPORTANT: \label is *after* \caption
\end{figure}

\begin{figure}[hbt]
  \begin{center}
    \resizebox{!}{60mm}{\includegraphics{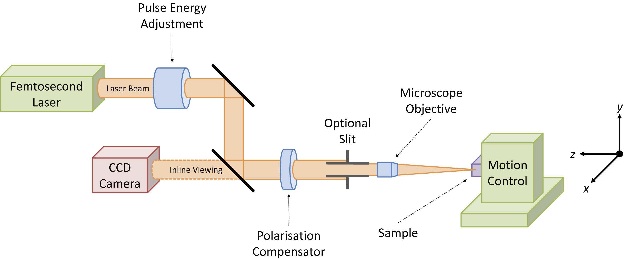}}
  \end{center}
  \caption{A diagram of the setup of the direct-write fabrication procedure. (Provided by M. Ams.)}
  \label{writesch}   % <-- IMPORTANT: \label is *after* \caption
\end{figure}

\begin{figure}[hbt]
  \begin{center}
    \resizebox{!}{60mm}{\includegraphics{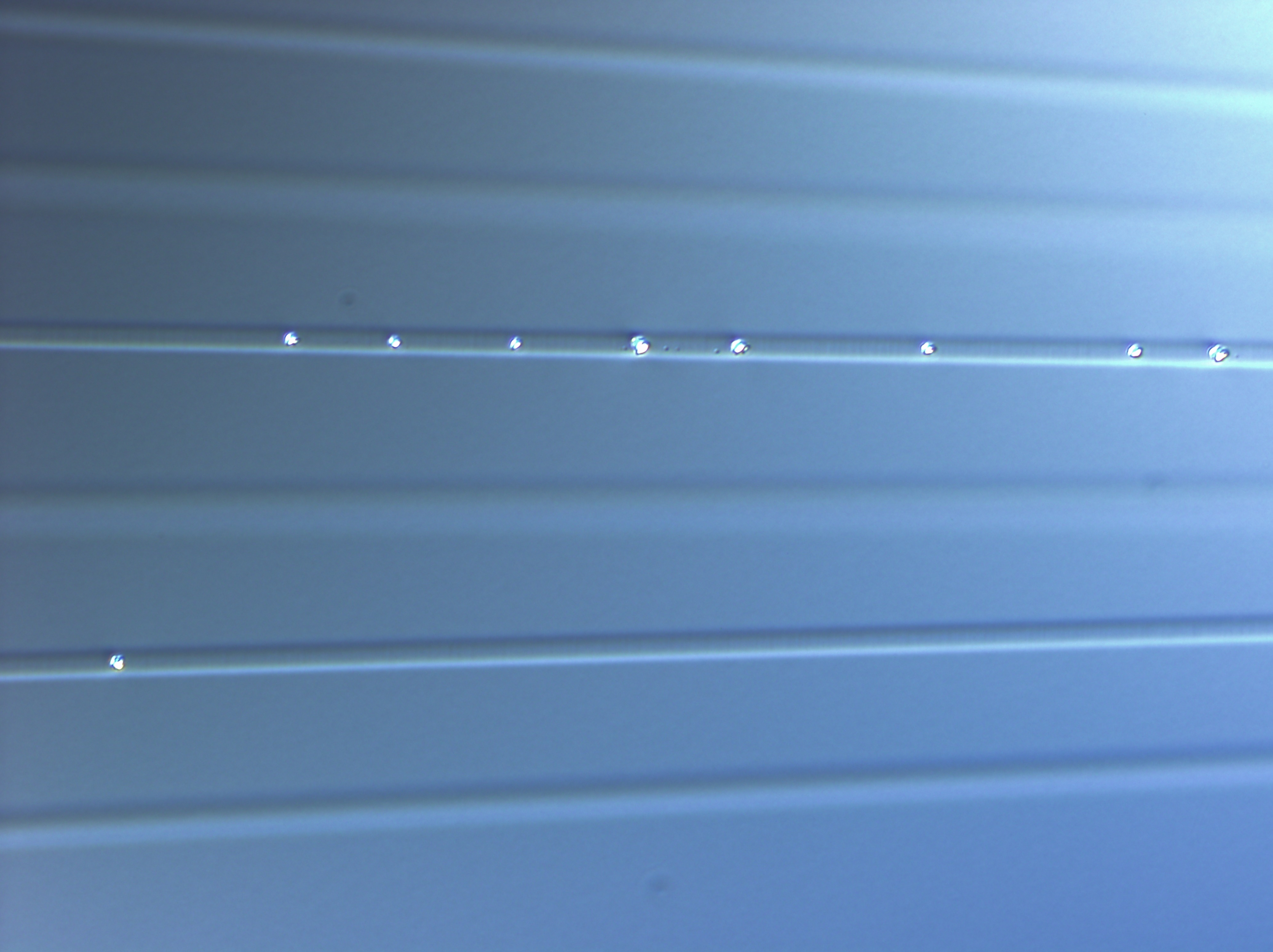}}
  \end{center}
  \caption{The fan-in section of the failed eight-pass tube fabrication. (Provided by G. D. Marshall.)}
  \label{damaged}   % <-- IMPORTANT: \label is *after* \caption
\end{figure}

\begin{figure}[hbt]
  \begin{center}
    \resizebox{!}{60mm}{\includegraphics{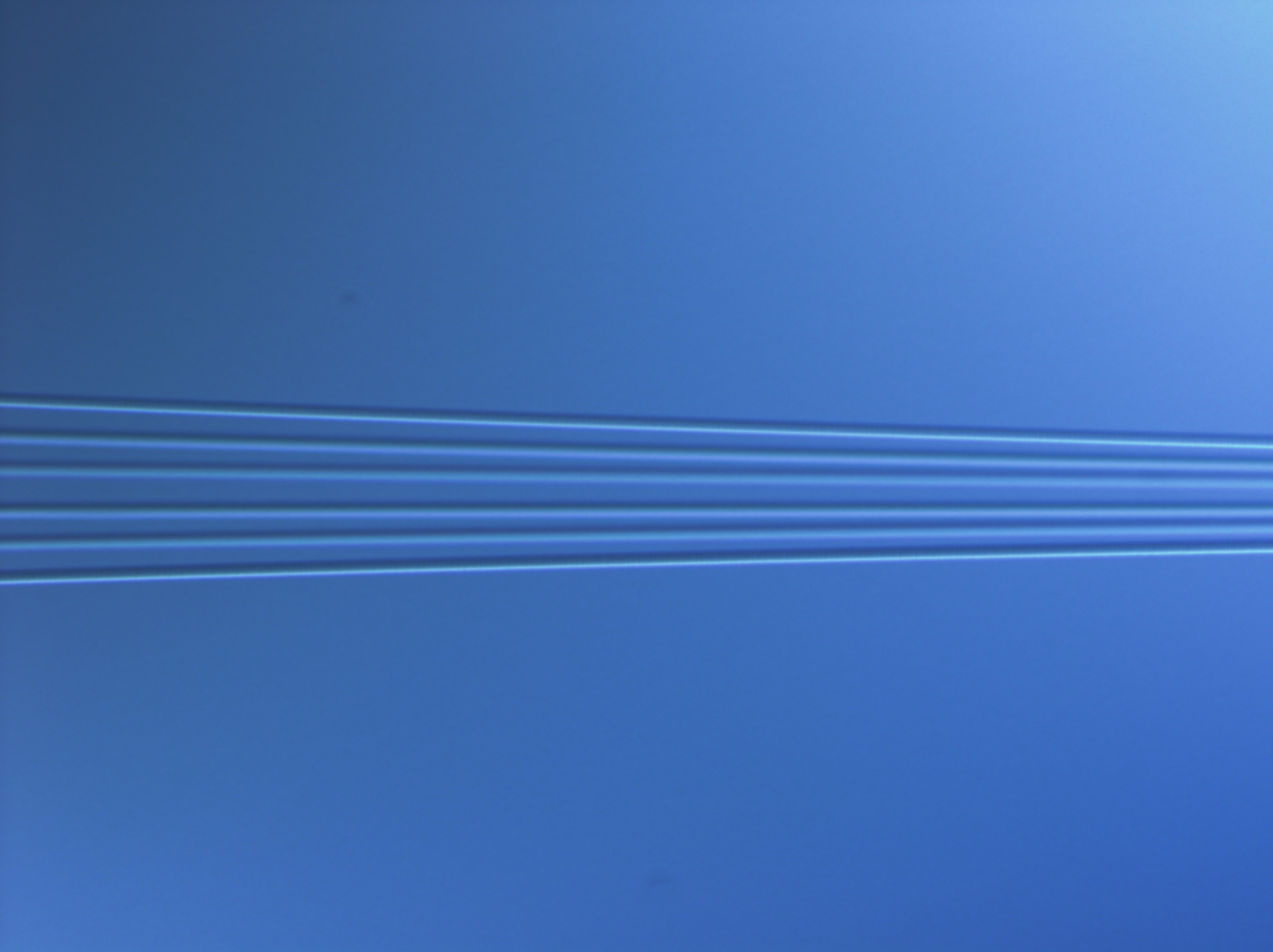}}
  \end{center}
  \caption{The fan-in section of the successful sixteen-pass tube fabrication. (Provided by G. D. Marshall.)}
  \label{16p1s}   % <-- IMPORTANT: \label is *after* \caption
\end{figure}

\begin{figure}[hbt]
  \begin{center}
    \resizebox{!}{60mm}{\includegraphics{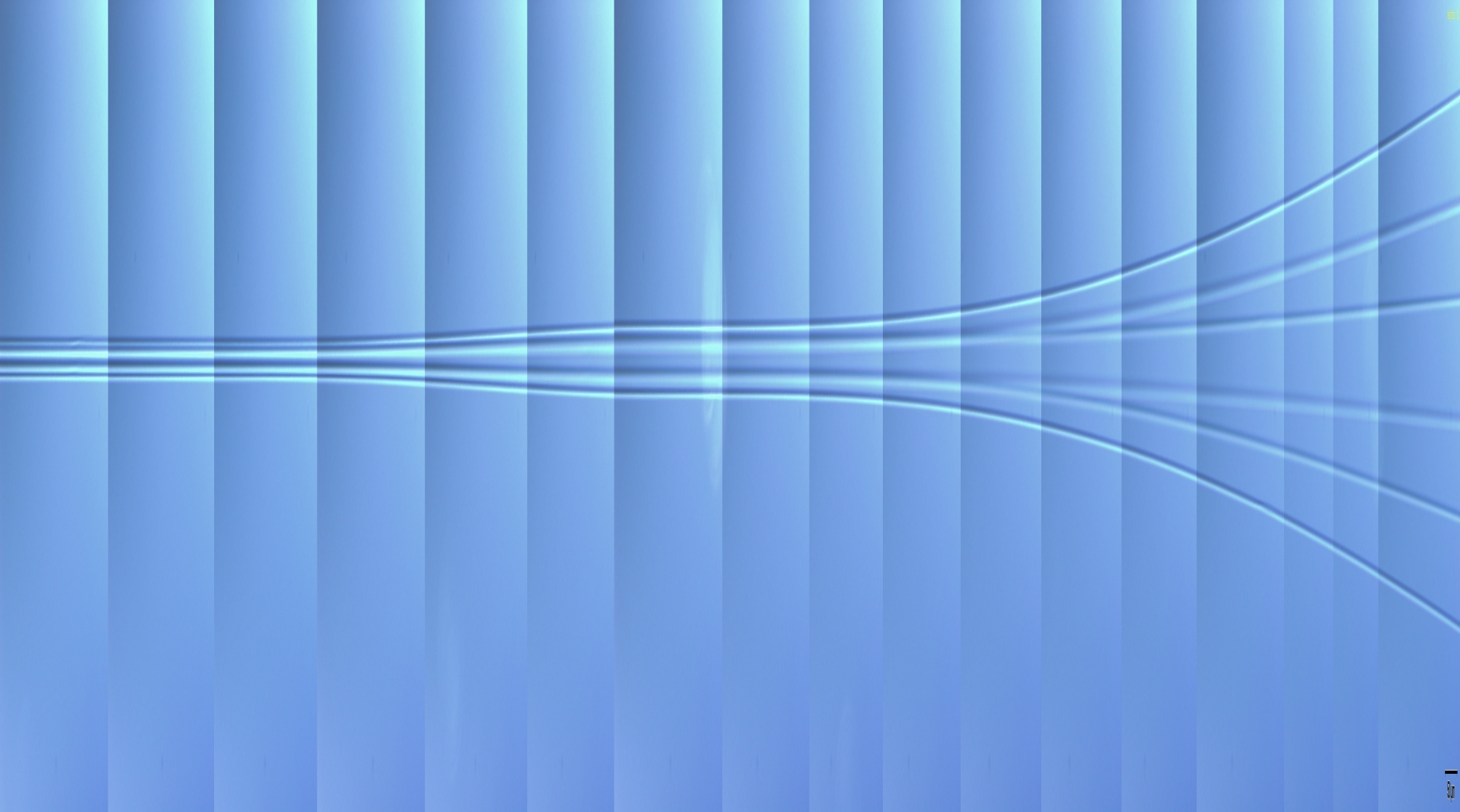}}
  \end{center}
  \caption{A montage combining different sections of the successfully fabricated two-stage tube. The aspect ratio of this figure is compressed by a factor of ten in the horizontal direction. The two stages of tube can be clearly seen with the intermediate tube occuring after the first stage of the fan-in section. (Provided by G. D. Marshall.)}
  \label{8p2s}   % <-- IMPORTANT: \label is *after* \caption
\end{figure}

The setup of the laser direct-write fabrication laboratory is shown in Fig.~\ref{writepic} and Fig.~\ref{writesch}. After fabrication, the devices are examined with an Olympus IX 81 transmission differential interference contrast microscope. A selection of images from this examination are shown as Fig.~\ref{damaged}, Fig.~\ref{16p1s} and Fig.~\ref{8p2s}.

The eight-pass tube which failed because of power-fluctuations in the laser during fabrication is shown in Fig.~\ref{damaged}, we can see that the waveguides are filled with white ``bubbles'' where scattering voids have been created. 

By comparison, in Fig.~\ref{16p1s} we see the result of a successful fabrication of the single-stage tube. We can note that the all the waveguides are very sharply defined here, whereas the failed fabrication shows that except in the waveguides which were subject to the power fluctuations, as evidenced by the white bubbles, waveguides created with eight passes have a weaker refractive contrast than the sixteen-pass waveguides.

In Fig.~\ref{8p2s}, sections of contrast microscopy photos of the eight-pass two-stage tube are shown to give a (fore)shortened impression of the tube design.

\section{Measurements}

\begin{figure}[hbt]
  \begin{center}
    \resizebox{!}{60mm}{\includegraphics{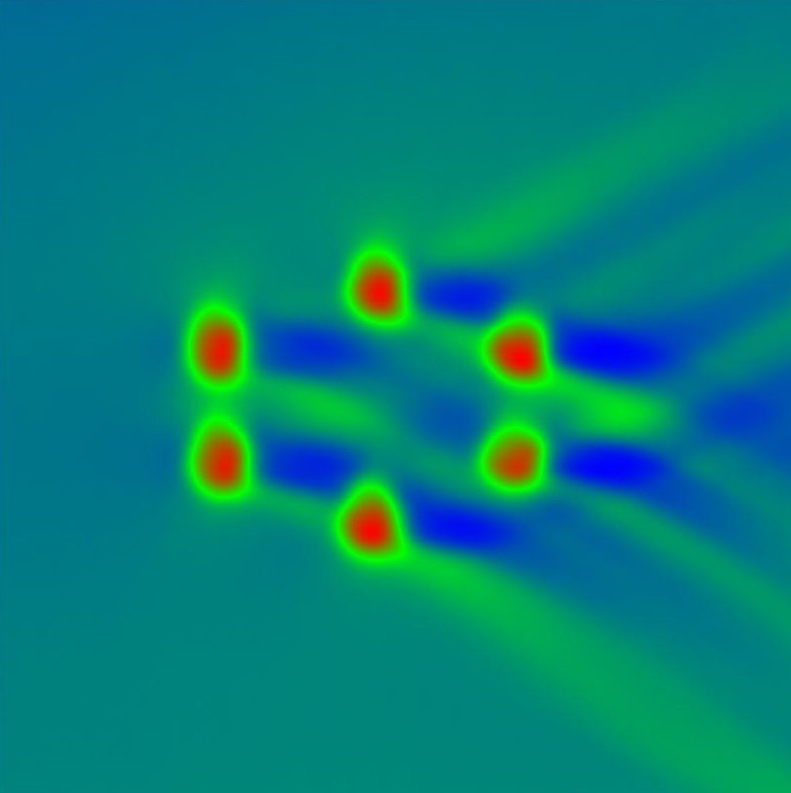}}
  \end{center}
  \caption{A screenshot from refractive index profilometry of the 16 pass tube. A qualitative linear contour plot of the 16-pass hexagonal tube region. The ``shadows'' of the waveguide structures extending to the right hand side are an artifact of the measurement instrument. In this figure the surface of the sample (the face through which the waveguides were written) is on the left. (Provided by G. D. Marshall.)}
  \label{16rip}   % <-- IMPORTANT: \label is *after* \caption
\end{figure}

The refractive index profile was measured with refractive contrast profilometry using Rinck elektronik equipment. The calibration is pending, but the image in Fig.~\ref{16rip} shows that the 16-pass tube corresponds to the mode profiles used in this project.

After the monolithic fused silica slide has been polished, bright classical laser light at approximately 780 nm is injected with a bulk optical fiber into a single waveguide in each of the two well-formed tubes. Using a Spiricon CCD camera with a microscope objective, beam profilometry is performed. The results are displayed graphically, in Fig.~{beam1} and Fig.~{beam2}. We can clearly see that the relationship between the beam profiles produced by injecting bright classical light into one specific waveguide is mirrored by the beam profile produced by injecting the light into the waveguide which lies opposite in the tube. In other words, the beam profile produced by injecting light into waveguide $1$ mirrored the one produced by injecting light into waveguide $6$, and the waveguide pairs $2$,$5$, and $3$,$4$ also mirror each other. The waveguide numbering convention used here has been illustrated in Fig.~\ref{6wg2stageRsoft}. \\

\begin{figure}[hbt]
  \begin{center}
    \resizebox{!}{70mm}{\includegraphics{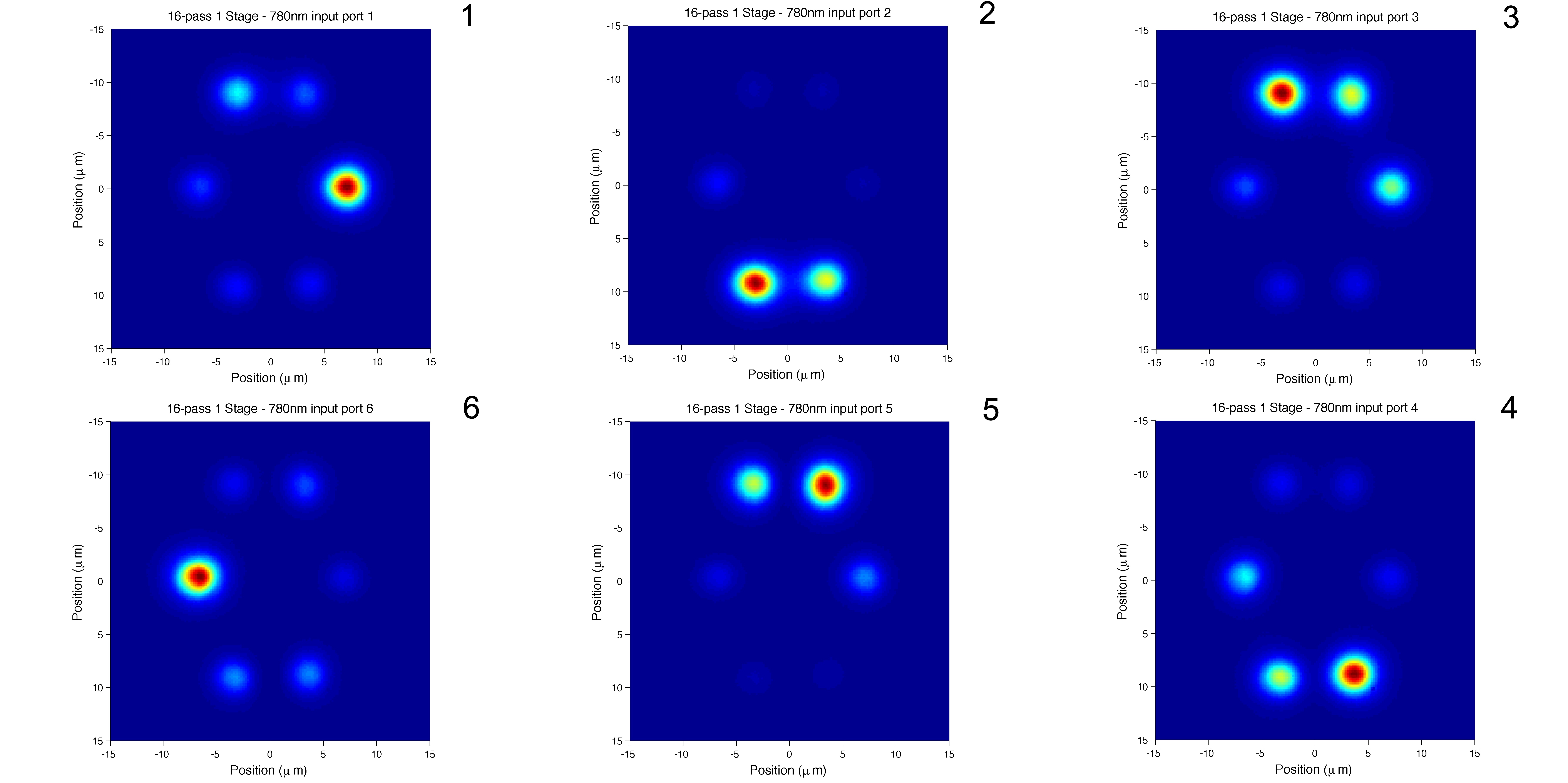}}
  \end{center}
  \caption{The beam profiles from injecting light into the single-stage tube, in waveguides $1-6$ by the numbering convention illustrated in Fig.~\ref{6wg2stageRsoft}. (Provided by G. D. Marshall.)}
  \label{beam1}   % <-- IMPORTANT: \label is *after* \caption
\end{figure}

\begin{figure}[hbt]
  \begin{center}
    \resizebox{!}{70mm}{\includegraphics{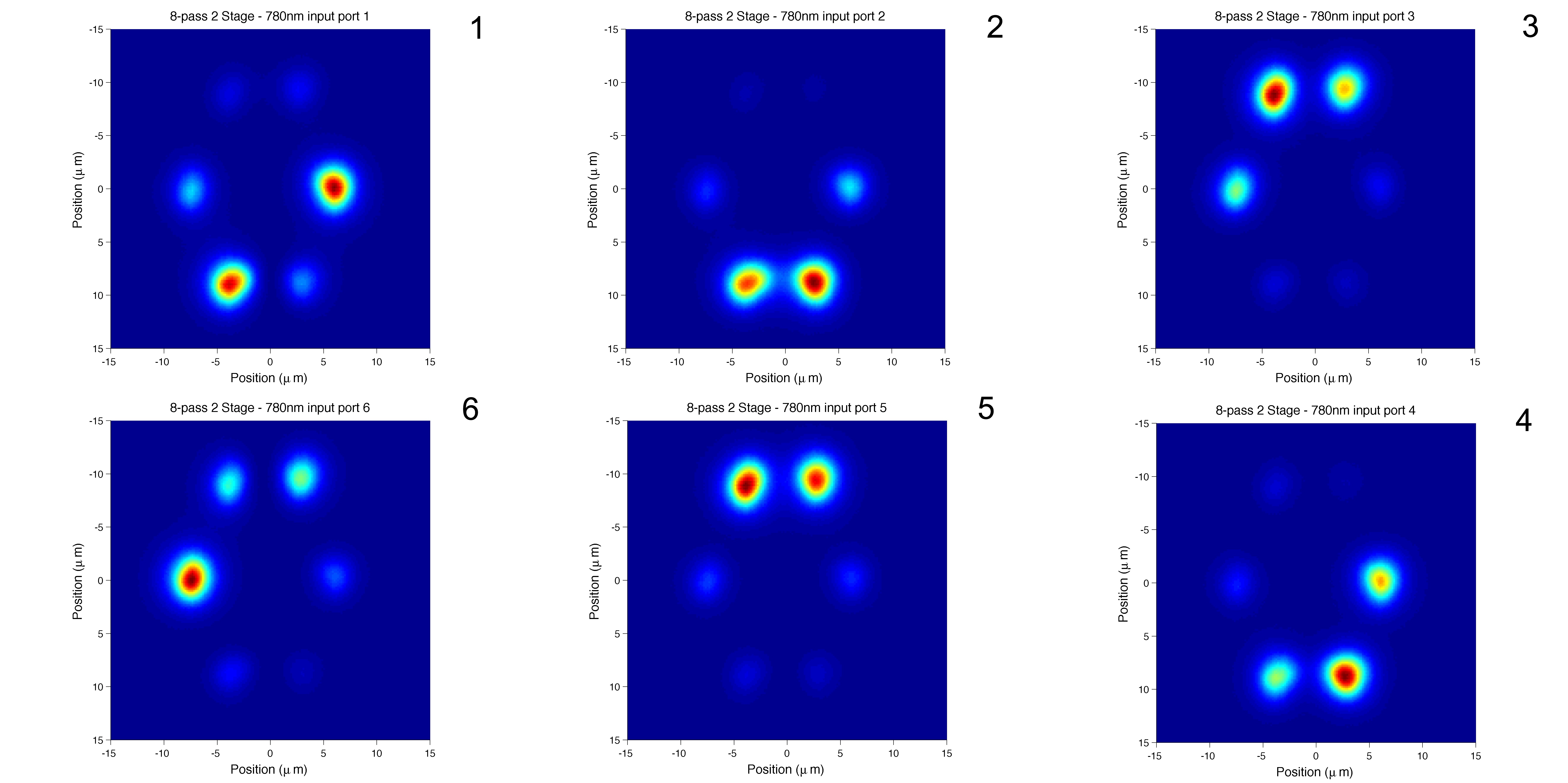}}
  \end{center}
  \caption{The beam profiles from injecting light into the two-stage tube, in waveguides $1-6$ by the numbering convention illustrated in Fig.~\ref{6wg2stageRsoft}. (Provided by G. D. Marshall.)}
  \label{beam2}   % <-- IMPORTANT: \label is *after* \caption
\end{figure}

Unfortunately, it has not been possible to directly study the QRW on the tubular waveguide arrays produced in this project. A suitable source of degenerate photon pairs is under construction at Macquarie University, but it was not available to this project. However, the successful design and fabrication of the tubular waveguide arrays has been verified by bright classical light. The hexagonal cross-section of the tubular waveguide array has an obvious symmetry, and each waveguide has a mirroring waveguide. For example, injecting light in waveguide 2 produces a coupling pattern which is the mirror image of the observed output coupling pattern that arises when light is injected into waveguide 5, which mirrors waveguide 2.   Fig.~\ref{beam1} and Fig.~\ref{beam2} shows this property in both the 8 pass and 16 pass tubular arrays. \\

The waveguides in the tubular arrays display excellent transmission properties, which in turn indicates that the fabricated devices successfully couple the injected light through the fan-in sections and into the tube region intended for quantum mechanical interaction in the QRW. The symmetric coupling output patterns reflect that the devices are symmetric throughout their lengths and that the coupling that occurs between the waveguides is also symmetric. The hexagonal cross-section appears somewhat stretched vertically, which is not ideal and the reason for which is unknown, but the symmetry of the output patterns appears unaffected by the slight stretch. \\

\section{Summary}
The experimental side of the project has been very successful in realising the designs the have been developed before and during the device fabrication. The two devices which have been produced seem to function well in the simple classical light tests they have so far been subjected to. It is interesting to note the slight vertical elongation of the tubular array cross-section, the reason for which remains unclear.

%**********************************************************************
%
%												Chapter 5 - Discussion
%
%**********************************************************************

\chapter{Discussion}

The project has consisted of theoretical as well as experimental developments. Initially, the basic mechanism of coupling was modelled and investigated, finding the relationship between waveguide separation and the length of propagation required for complete coupling to occur. Then, the classical experimental results reported by Bromberg {\it et al.} were simulated using RSoft set at parameters which the direct-write technique can fabricate, reproducing the overall pattern in the correlation matrices. Various alternative configurations of the tubular waveguide array have been investigated in RSoft and with Coupled-Mode Theory code in Mathematica. Some investigations have also been performed with unrealistic values for certain physical parameters. With a large refractive contrast or waveguide widths, the discrepancy between RSoft and CMT was reduced. The physical constraints created by non-nearest-neighbour coupling (NNNC) have been identified. In response to this challenge, a functional compromise of variables has been found without radically re-designing the device. \\

In the first fabrication run, the production of two devices was intended but one failed. However, the succesfully produced device was the more innovative. With the sixteen pass device, there is a hope that the refractive contrast limit of 0.00455 can be overcome, if only fractionally. The second production run created a two-stage device in eight passes. Both devices work as expected, and the current beam profilometry measurements of the two devices are extraordinarily encouraging, mainly because the beam profiles are symmetric, as illustrated in Fig.~\ref{beam1} and Fig.~\ref{beam2}. This for the first time demonstrates that three-dimensional coupling networks can now be built. This is potentially very promising results for the field of quantum optics and supports the idea that optical implementations of quantum computing may be the most practical alternative. Furthermore, this development provides the opportunity to further study the fundamental properties of quantum mechanical systems. \\

From the beam profilometry, we can surmise that the two-stage fan-in appears to give the more symmetrical set of beam profiles. However, a final conclusion on this design issue depends on actual quantum photonics experiments with single photon sources, to be performed in the future. Ultimately, the current measurements can not determine with absolute certainty if the degenerate single-photon case will behave as quantum mechanics predicts. Another factor in deciding which of the two devices is the best, may be further simulation. Primarily, we see that this novel application of the direct-write technique has resulted in functional waveguides in a three-dimensional array. We also note that some of the waveguides have been direct-written through pre-existing waveguides, and while this had not previously been attempted and there was concern that refraction from one waveguide could possibly perturb the writing of another, the devices appear well-formed. \\

\section{Future outlook}

\begin{figure}[hbt]
  \begin{center}
    \resizebox{!}{70mm}{\includegraphics{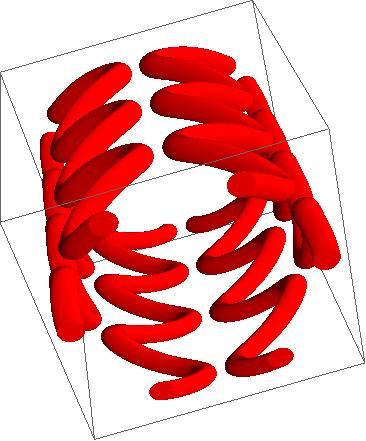}}
  \end{center}
  \caption{The possible future design of the wriggler, by J. Twamley. (Provided by J. Twamley.)}
  \label{wriggler}   % <-- IMPORTANT: \label is *after* \caption
\end{figure}

An entirely different design is also being considered, for future investigations. Twamley is developing an alternative tubular structure called a wriggler, which is shown in Fig.~\ref{wriggler} to perform QRWs. The device is called a wriggler and it avoids non-nearest neighbour coupling by being a tube of waveguides with a greater tube radius than the devices produced in this project. The waveguides in the wriggler are not always parallel, but wriggle back and forth on the circumference of the tube, with coupling ideally only occurring in sections where two waveguides are adjacent and parallel. However, this is not part of the work presented here. We can simply note that while the tubular array of parallel waveguides investigated so far can implement a continuous QRW, the wriggler would implement a discrete QRW.  \\

Mainly due to non-nearest-neighbour coupling in simulations, it seems that the tubular waveguide array devices in their current state may not be the most apt for performing a simple quantum random walk on a circle. The implication of the simulation results is that the tubular array of parallel waveguides fabricated here is not the most appropriate design for performing QRWs in a waveguide network analogous to a walk on a circle. Instead, the wriggler design seems to be a more promising direction for the next step in investigating this QRW on a circle, specifically. Also, that design makes the quantum walk discrete as opposed to continuous. However, if the waveguides are contiguous and well-fabricated enough to couple degenerate photons without causing decoherence, then on the one hand, the geometry of the QRW in the tubular waveguide array may not be so simple. On the other hand, the devices produced can still provide a unique opportunity for investigating photonic quantum interference.

\section{Conclusion}

The project has succeeded in creating novel three-dimensional devices in integrated photonics for future quantum optics studies of quantum random walks. This represents a development not only for quantum random walks, but potentially also for any optical quantum processing applications where a compact three-dimensional architecture can provide an advantage.
This project is a key technological demonstration of the theory and practice behind creating a three-dimensional linear optical circuit implementation of continuous array QRWs. It has opened the door to enabling the study of this field by making use of the novel three-dimensional waveguide circuit manufacturing capabilities offered by the direct-write technique. 
\\

\bibliographystyle{bibthesis}

\bibliography{LitRev2}
\end{document}